\PassOptionsToPackage{table}{xcolor}
\documentclass[leqno,oneeqnum,onefignum,onetabnum,final]{siamart190516}

\usepackage[T1]{fontenc}
\usepackage[plainruled,noend,linesnumbered]{algorithm2e}
\usepackage{amsmath,amssymb,amsfonts,amsbsy,amsfonts,latexsym}
\usepackage{array}
\usepackage{multirow}
\usepackage{makecell}
\usepackage{subcaption}
\usepackage[labelfont=bf,textfont=it,belowskip=0pt,aboveskip=5pt,tableposition=top]{caption}
\usepackage[noadjust]{cite}
\usepackage{xcolor}
\usepackage{colortbl}
\usepackage{siunitx}
\usepackage{booktabs}
\usepackage{times}
\usepackage{nicefrac}
\usepackage{arydshln}
\usepackage{graphicx}
\usepackage{rotating}
\usepackage{placeins}
\usepackage{fancyhdr}
\usepackage{xspace}
\newcounter{runidnum}
\usepackage{geometry}
\definecolor{colorA}{RGB}{189,201,225}
\definecolor{colorB}{RGB}{103,169,207}
\definecolor{colorC}{RGB}{ 28,144,153}
\definecolor{colorD}{RGB}{  1,108, 89}

\newcolumntype{R}{>{\columncolor{gray!40}}r}
\newcolumntype{L}{>{\columncolor{gray!40}}l}
\newcolumntype{C}{>{\columncolor{gray!40}}c}

\graphicspath{{./figs/}}
\DeclareGraphicsExtensions{.pdf,.png}

\usepackage{pgfplots, pgfplotstable}

\usepackage{paralist}

\let\pgfimageWithoutPath\pgfimage
\renewcommand{\pgfimage}[2][]{\pgfimageWithoutPath[#1]{figs/#2}}

\DeclareCaptionLabelFormat{andtable}{#1~#2  \&  \tablename~\thetable}

\makeatletter
\def\@algocf@capt@plainruled{above}
\renewcommand{\algocf@caption@plainruled}{%
  \vskip\AlCapSkip%
  \box\algocf@capbox%
  \vskip 5\algoheightrule}%
\makeatother

\usepackage{xparse}% http://ctan.org/pkg/xparse
\NewDocumentCommand{\var}{O{s} m O{}}{%
  \ensuremath{#1_{#2}^{#3}}% add \vphantom{<bizarre sup>}
}

\newcommand\revise[1]{#1}

\newcommand{\pcIR}{Inv$\mathcal{A}$\xspace}        % short cut for spectral preconditioner
\newcommand{\pcIHN}{Inv$\mathcal{H}_0$\xspace}     % short cut for inverse H0 preconditioner
\newcommand{\pcIHNTL}{2LInv$\mathcal{H}_0$\xspace} % short cut for inverse H0 two level preconditioner

\newcommand{\figref}[1]{Figure~\ref{#1}}
\newcommand{\tabref}[1]{Table~\ref{#1}}
\newcommand{\secref}[1]{\S\ref{#1}}
\newcommand{\algref}[1]{Algorithm~\ref{#1}}

\newcommand{\vect}[1]{\boldsymbol{#1}} % vector
\newcommand{\defeq}{\ensuremath{\mathrel{\mathop:}=}}

\newcommand{\iquote}[1]{``\textit{#1}''}
\newcommand{\acr}[1]{\textbf{#1}}

\newcommand{\sw}[1]{\texttt{#1}}
\newcommand{\claire}{\texttt{CLAIRE}}
\newcommand{\lddmm}{LDDMM}

\DeclareMathOperator*{\minopt}{minimize}

\newcommand{\ipoint}[1]{\textit{\textbf{\color{darkgray}#1}}}

\newcommand{\ns}[1]{\ensuremath{\mathbb{#1}}}
\newcommand{\fun}[1]{\ensuremath{\mathcal{#1}}}
\newcommand{\idiv}{\ensuremath{\nabla\cdot}}
\newcommand{\igrad}{\ensuremath{\nabla}}

\newcommand{\half}[1]{\frac{#1}{2}}
\newcommand{\T}{\ensuremath{\mathsf{T}}}

\newcommand{\dop}[1]{\ensuremath{\mathcal{#1}}}
\newcommand{\di}[1]{\ensuremath{\mathbf{#1}}}

\newcommand\x{\ensuremath{\vect{x}}}
\newcommand\sta{\ensuremath{m}}
\newcommand\vel{\ensuremath{\vect{v}}}
\newcommand\dmap{\ensuremath{\vect{y}}}
\newcommand\adj{\ensuremath{\lambda}}
\newcommand\iadj{\ensuremath{\tilde{\lambda}}}
\newcommand\ista{\ensuremath{\tilde{m}}}
\newcommand\ivel{\ensuremath{\vect{\tilde{v}}}}
\newcommand\traj{\ensuremath{\vect{y}}}

\newcommand\divel{\ensuremath{\di{\tilde{v}}}}
\newcommand\dvel{\ensuremath{\di{v}}}

\renewcommand{\d}[1]{\mathop{}\!\mathrm{d}#1}
\newcommand{\dt}{\d{t}}
\newcommand{\dx}{\d{\x}}
\newcommand{\p}{\partial}

\def\algcolor#1{\leavevmode\rlap{\hbox to \hsize{\color{#1!30}\leaders\hrule height .8\baselineskip depth .5ex\hfill}}}
\def\algcomment#1{{\hfill$\triangleright$\small\textsf{ #1 }}}

\newcommand{\bipa}{\begin{inparaenum}[(\itshape i\upshape)]}
\newcommand{\eipa}{\end{inparaenum}}
\newcommand{\bipasub}{\begin{inparaenum}[(\itshape a\upshape)]}
\newcommand{\eipasub}{\end{inparaenum}}

\definecolor{light-gray}{gray}{0.80}

\newcommand{\approxspeed}[1]{\textasciitilde$#1$\texttt{x}}

\newcommand{\PLH}{{\mkern-2mu\times\mkern-2mu}}

\renewcommand\paragraph{\subsubsection*}

\title{Multi-Node Multi-GPU Diffeomorphic Image Registration for Large-Scale Imaging Problems\thanks{This work was partly supported by the National Science Foundation (DMS-1854853, DMS-2009923, DMS-2012825, CCF-1817048, CCF-1725743), the NVIDIA Corporation (NVIDIA GPU Grant Program), the Deutsche Forschungsgemeinschaft (DFG, German Research Foundation) under Germany's Excellence Strategy-EXC 2075-390740016,  by the U.S. Department of Energy, Office of Science, Office of Advanced Scientific Computing Research, Applied Mathematics program under Award Number DE-SC0019393; by the U.S. Air Force Office of Scientific Research award FA9550-17-1-0190;  by the Portugal Foundation for Science and Technology and the UT Austin-Portugal program, and by NIH award 5R01NS042645-11A1. Any opinions, findings, and conclusions or recommendations expressed herein are those of the authors and do not necessarily reflect the views of the DFG, AFOSR, DOE, NIH, and NSF. Computing time on the Texas Advanced Computing Centers' (TACC) systems was provided by an allocation from TACC and the NSF. This work was completed in part with resources provided by the Research Computing Data Core at the University of Houston.}}

\author{
Malte Brunn\footnotemark[1], Naveen Himthani\footnotemark[2], George Biros\footnotemark[2], Miriam Mehl\footnotemark[2] \and Andreas Mang\footnotemark[3]
}

\begin{document}
\maketitle

\renewcommand{\thefootnote}{\fnsymbol{footnote}}

\footnotetext[1]{Computer Science, University of Stuttgart, Stuttgart, DE, Email: \{{\tt malte.brunn,miriam.mehl}\}{\tt @ipvs.uni-stuttgart.de}}
\footnotetext[2]{Oden Institute, University of Texas, Austin TX, US, Email: \{{\tt naveen@oden.utexas.edu}, {\tt gbiros@acm.org}\}}
\footnotetext[3]{Mathematics, University of Houston, Houston TX, US, Email: {\tt andreas@math.uh.edu}}

%//////////////////////////////////////////////////////////////
\begin{abstract}
We present a  Gauss-Newton-Krylov solver for large deformation diffeomorphic image registration. We extend the publicly available  CLAIRE library to  multi-node multi-graphics processing unit (GPUs) systems and introduce novel algorithmic modifications that significantly improve  performance.  Our contributions comprise \bipa\item a new preconditioner for the reduced-space Gauss-Newton Hessian system,  \item a highly-optimized multi-node multi-GPU implementation exploiting device direct communication  for the main computational kernels (interpolation, high-order finite difference operators and Fast-Fourier-Transform), and \item a comparison with state-of-the-art CPU and GPU implementations\eipa. We solve a  $256^3$-resolution image registration problem in five seconds on a single NVIDIA Tesla V100, with a performance speedup of 70\% compared to the state-of-the-art. In our largest run, we register $2048^3$ resolution images (25\,B unknowns; approximately 152$\times$ larger than the largest problem solved in state-of-the-art GPU implementations) on 64 nodes with 256 GPUs on TACC's Longhorn system.
\end{abstract}
%//////////////////////////////////////////////////////////////

\section{Introduction}\label{s:intro}

\begin{figure*}
\centering
\includegraphics[width=0.95\textwidth]{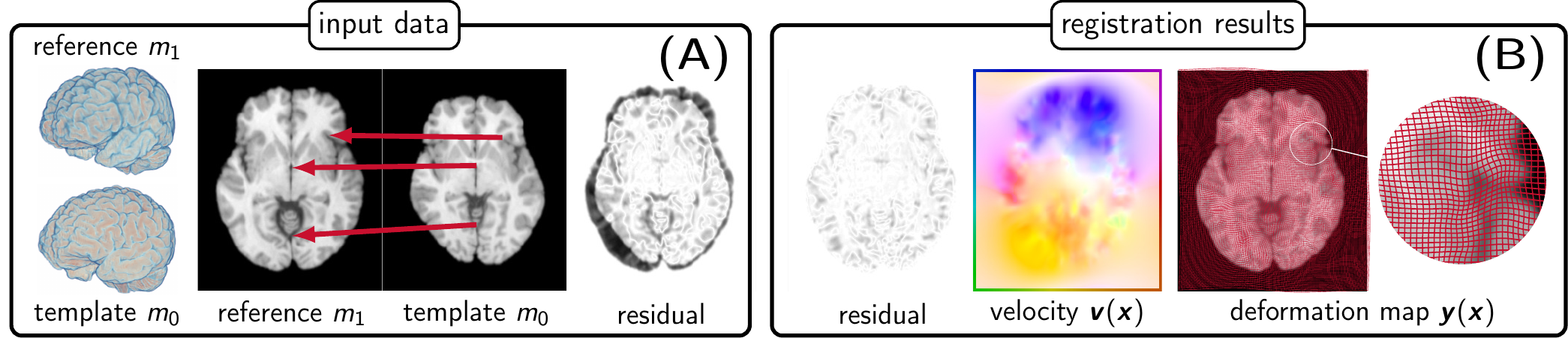}
\caption{3D diffeomorphic image registration problem for human neuroimaging data. We illustrate the input data and the registration problem in panel (A) and the results in panel (B) for a multi-subject registration problem (NIREP dataset~\cite{Christensen:2006a}; reference image: \texttt{na01}; template image: \texttt{na10}). Panel (A) [from left to right]: Volume rendering of the input images, axial view of the reference image, axial view of the template image, and the residual of these views before registration (white: small residual; black: large residual). The image registration problem is to identify spatial correspondences that map points from one image (the template image) to points in another image (the reference image); see red arrows. Panel (B) [from left to right]: residual after registration, computed velocity field $\vel(\x)$ that parameterizes the deformation map $\dmap(\x)$ (color denotes orientation), and an illustration of $\dmap(\x)$. Qualitatively, the computed map is a smooth diffeomorphism (confirmed numerically).\label{f:regprob-brains}}
\end{figure*}

\begin{figure*}
\centering
\includegraphics[width=0.95\textwidth]{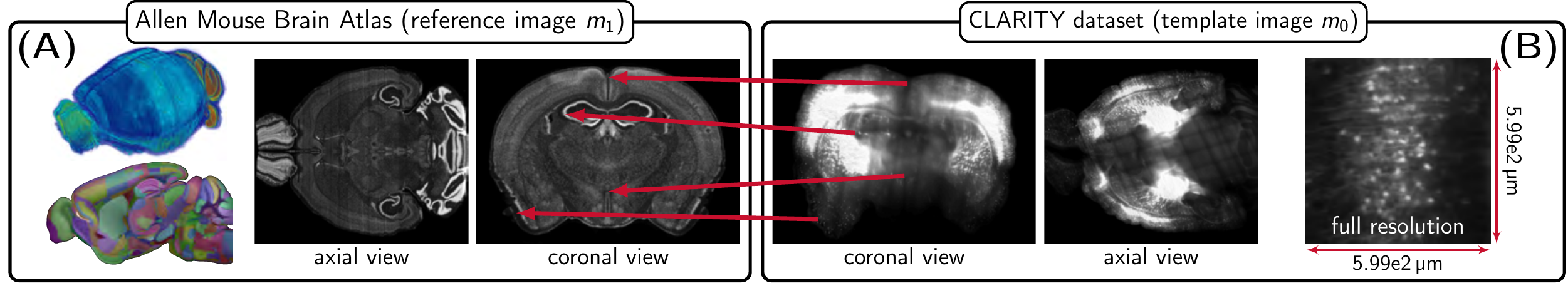}
\caption{3D image registration problem for murine CLARITY imaging data. We illustrate a multi-subject registration problem. In panel (A), we show the Allen Mouse Brain Atlas~\cite{Jones:2009a} with a grid size of $800\times1140\times1320$. We show a volume rendering (top left), annotations of anatomical regions (bottom left), an axial view (middle) and a coronal view (right). In panel (B), we show a coronal and an axial view (after affine pre-registration to the atlas image), as well as a closeup of a subregion in full resolution. This CLARITY volume (Control 189) has a resolution of $\SI{585}{\nano\metre} \times \SI{585}{\nano\metre} \times \SI{5}{\micro\metre}$ with a grid size of $\num{20084}\times\num{24618}\times\num{1333}$. Once we have found the diffeomorphism, we can transfer the annotations of the anatomical regions identified in the atlas (see panel (A)) to the CLARITY dataset, and study anatomical subregions.\label{f:regprob-clarity}}
\end{figure*}

3D diffeomorphic image registration is a critical task in biomedical imaging applications~\cite{Fischer:2008a,Modersitzki:2004a,Sotiras:2013a}. For example, it enables the analysis and study of morphological changes associated with the progression of neurodegenerative diseases in time series of medical images or in imaging studies of patient populations. The input to this inverse problem are two (or more) images $\sta_0(\x)$ (the \iquote{template image}) and $\sta_1(\x)$ (the \iquote{reference image}) of the same type of object, compactly supported on a domain $\Omega \subset\ns{R}^3$. The task of image registration is to compute a spatial transformation or mapping $\dmap(\x)$ such that $\sta_0(\dmap(\x)) \approx \sta_1(\x)$ for all points $\x\in\Omega$~\cite{Modersitzki:2004a} (see~\figref{f:regprob-brains}). Methods for the registration of images can be classified according to the parameterization for $\dmap$~\cite{Modersitzki:2004a}. We will consider maps $\dmap$ that are \ipoint{diffeomorphisms}, i.e., maps that are a differentiable bijection, and have a differentiable inverse. In the present work, we consider formulations that belong or are related to a class of methods referred to as \emph{large-deformation diffeomorphic metric mapping} (\acr{LDDMM})~\cite{Beg:2005a,Trouve:1998a,Younes:2010a}. These methods parameterize diffeomorphisms in terms of a smooth (time-dependent) velocity field. The associated mappings provide maximal flexibility~\cite{Sotiras:2013a} but are expensive to compute: the problem is infinite-dimensional, and upon discretization it becomes a nonlinear system with millions or even billions of unknowns. For example, registering two volumes of grid size $256^3$ (a typical data size for clinical images) necessitates solving for approximately 50\,M unknowns (three vector components per image grid point). This is further complicated by the fact that image registration is a highly non-linear, ill-posed inverse problem~\cite{Fischer:2008a}, resulting in ill-conditioned inversion operators. As a result, image registration can take several minutes on multi-core high-end CPUs. As clinical workflows for multi-center population-studies that require thousands of registrations become increasingly more common, execution time of a single registration becomes more and more critical; reducing the runtime to seconds corresponds to a reduction of clinical study time from weeks to a few days. GPUs with their inherent parallelism and low energy consumption are an attractive choice to achieve this goal. However, despite the need for high computational throughput and the existence of several software packages for \lddmm{}, there is little work on high-performance GPU implementations, and even less work on multi-node multi-GPU implementations for large-scale applications. One such application is the registration of CLARITY images~\cite{Chung:2013b,Kim:2013a,Kutten:2016a,Kutten:2017a,Tomer:2014a,Vogelstein:2018a} of resolution in the order of $20\text{\,K}\times20\text{\,K}\times1\text{\,K}$, which corresponds to a problem with about 1.2 trillion unknowns (see~\figref{f:regprob-clarity}).

\begin{table}
\caption{Notation and main symbols.\label{t:notation-and-symbols}}
\centering\scriptsize
\begin{tabular}[t]{lllll}\toprule
Symbol                  & Description                                                                      \\\midrule
$\Omega$                & spatial domain; $\Omega\defeq[0,2\pi)^3\subset\ns{R}^3$ with boundary $\p\Omega$ \\
$\x$                    & spatial coordinate; $\vect{x}\defeq(x_1,x_2,x_3)^\T\in\ns{R}^3$                  \\
$t$                     & pseudo-time variable; $t \in [0,1]$                                              \\
$\sta_1(\x)$            & reference image                                                                  \\
$\sta_0(\x)$            & template image (image to be registered)                                          \\
$\vel(\x)$              & stationary velocity field                                                        \\
$\vect{y}(\x)$          & deformation map                                                                  \\
$\sta(\vect{x},t)$      & state variable (transported intensities of $m_0$)                                \\
$\lambda(\vect{x},t)$   & adjoint variable                                                                 \\
$\dop{A}$               & regularization operator                                                          \\
$\beta > 0$             & regularization parameter                                                         \\
\bottomrule
\end{tabular}
\end{table}

\subsection{Formulation and Outline of the Method}\label{s:form}

We summarize our notation in \tabref{t:notation-and-symbols}. \claire{} uses an optimal control formulation. The deformation map $\dmap(\x)$ is parameterized through a smooth, stationary velocity field $\vel(\x)$. The optimization problem is: given two images $\sta_0(\x)$ (template image; image to be deformed) and $\sta_1(\x)$ (reference image), we seek $\vel(\x)$ by solving
\begin{subequations}
\label{e:problem}
\begin{equation}
\label{e:varopt:objective}
\minopt_{\vel, m}\;\;
\half{1}\!\int_{\Omega}\!(\sta(\x,1) - \sta_1(\x))^2\!\dx
+\frac{\beta}{2} \operatorname{reg}(\vel)
\end{equation}
\noindent subject to
\begin{align}
\p_t \sta(\x,t) + \vel(\x) \cdot\!\igrad \sta(\x,t) &= 0 &&\text{in}\;\Omega \times(0,1], \nonumber \\
\sta(\x,t) &= m_0(\x)                                    &&\text{in}\;\Omega \times\{0\}  \label{e:varopt:constraint}
\end{align}
\end{subequations}

\noindent on a three-dimensional rectangular domain $\Omega$ with periodic boundary conditions on $\p\Omega$. The first term in~\eqref{e:varopt:objective} is a similarity measure for the proximity between the deformed template image $m(\x,1)$ and the reference image $m_1(\x)$. Without loss of generality, we consider a squared $L^2$-distance. The second term in~\eqref{e:varopt:objective} is a Tikhonov regularization functional with regularization parameter $\beta > 0$. This regularization operator is not only introduced to alleviate issues with the ill-posedness of the inverse problem but also prescribes sufficient regularity requirements for $\vel(\x)$ to ensure that the computed geometric transformation is a diffeomorphism~\cite{Barbu:2016a, Borzi:2002a, Vialard:2012a, Beg:2005a, Chen:2011a, Younes:2010a}. The default configuration of \claire{} is an $H^1$-Sobolev-seminorm; the regularization model is a standard $L^2$-inner product $\langle\dop{A}\vel(\x),\vel(\x)\rangle_{L^2(\Omega)^3}$, where $\dop{A}$ is a vector Laplacian operator. The formulation is augmented with an additional penalty on the divergence of $\vel$ (see~\cite{Mang:2016H1,Mang:2018CLAIRE} for details). The transport equation~\eqref{e:varopt:constraint} describes the geometric transformation of the template image $\sta_0(\x)$ by advecting the intensities forward in time. We use a reduced-space Gauss--Newton--Krylov method to solve~\eqref{e:problem}. Details can be found in \secref{s:numerics}.

\subsection{Contributions \& Challenges}\label{s:contributions}
We extend the open source diffeomorphic image registration framework termed \claire{}~\cite{Brunn:2020a,Gholami:2017SC,Mang:2015NK,Mang:2016H1,Mang:2016SC,Mang:2018CLAIRE,claire-web}. \claire{} uses an optimal control formulation with partial differential equations (\acr{PDE}s; e.g., a pure advection equation for the image intensities) as constraints. The overall mathematical formulation and solution strategy has not been altered from~\cite{Mang:2018CLAIRE}. \claire{} has been developed to scale on standard x86 CPU clusters using the Message Passing Interface (\acr{MPI}) for parallelism~\cite{Gholami:2017SC,Mang:2016SC,Mang:2018CLAIRE,claire-web} and has recently been ported to GPU architectures (\ipoint{single-node single-GPU} implementation)~\cite{Brunn:2020a}.

In the present work, we propose a new, highly optimized multi-node multi-GPU implementation of \claire{}. The main \ipoint{challenges} are \bipa\item eliminating costly host-to-device copies, \item addressing significant communication costs between devices, \item reducing memory pressure to enable large-scale runs on limited resources, and \item identifying an adequate balance between parallelism and local computational throughput\eipa. (Per GPU we need to hold enough data to locally perform a sufficient amount of computations, since the computational kernels are extremely fast. Too few data to process per GPU deteriorates scalability. This effect is much more pronounced on GPUs compared to CPUs.) Our main \ipoint{contributions} are:
\begin{enumerate}
\item[\textbf{1)}]
We propose an efficient GPU-only single- and multi-node multi-GPU implementation of \claire{}. The proposed multi-GPU implementation is available for download at https://github.com/andreasmang/claire \cite{claire-web}.

\item[\textbf{2)}]
We minimize communication between host and device through CUDA-aware MPI, and increase the computational throughput in the most important computational kernels of the solver, scattered-data interpolation (\acr{IP}) and differentiation.

\item[\textbf{3)}]
We propose several improvements to reduce memory pressure and, thus, further increase the computational throughput. With the proposed implementation, it is possible to solve problems with datasets of grid sizes of $512^3$ on a single node using four NVIDIA Tesla V100 GPUs in under \SI{30}{\second}.

\item[\textbf{4)}]
We propose a completely new preconditioner for the reduced-space Hessian based on a zero-velocity approximation, which we term \iquote{\pcIHN}. This allows us to eliminate expensive incremental forward and adjoint PDE solves (hyperbolic transport equations) in the evaluation of the preconditioner. Our method is matrix-free (we do not store or assemble the preconditioner or the Hessian). To further amortize computational costs, we propose a two-level coarse grid approximation.

\item[\textbf{5)}]
We report results for synthetic and real data, which includes results for CLARITY imaging data for a grid size of $1024\PLH768\PLH768$. Overall, we achieve a speedup of up to about 70\% on a single GPU compared to the state-of-the-art~\cite{Brunn:2020a}. This makes the proposed solver $34\PLH$ faster than the CPU version~\cite{Mang:2016SC,Gholami:2017SC,Mang:2018CLAIRE} and $50\PLH$ faster than other, exemplary GPU-accelerated implementations for LDDMM (c.f., benchmark study in~\cite{Brunn:2020a}). Moreover, our multi-GPU implementation allows us to solve problems that are approximately 152$\times$ larger ($N = 2048^3$, 25\,B unknowns) compared to~\cite{Brunn:2020a}.
\end{enumerate}

\subsection{Limitations}\label{s:limitations}
We have optimized memory allocation for the core components of \claire{}. Additional optimizations by sharing memory across external libraries and parallel-in-time integration methods to further reduce the memory pressure remain subject to future work. Moreover, \claire{} uses stationary velocities. This drastically improves efficiency, but results in theoretical limitations.

\subsection{Related Work}\label{s:relwork}
The present work builds upon the open source framework termed \claire{}~\cite{Brunn:2020a,Gholami:2017SC,Mang:2015NK,Mang:2016H1,Mang:2016SC,Mang:2018CLAIRE,claire-web}. Related \lddmm{} software packages include \sw{Demons} \cite{Vercauteren:2009a}, \sw{ANTs} \cite{Avants:2011a,Avants:2008a,ants-web}, \sw{DARTEL} \cite{Ashburner:2007a}, \sw{deformetrica}~\cite{Bone:2018a,Bone:2018b,Fishbaugh:2017a,deformetrica-web}, \sw{FLASH} \cite{Zhang:2018a}, \sw{LDDMM} \cite{Beg:2005a,lddmm-web}, \sw{ARDENT} \cite{ardent-web}, \sw{ITKNDReg} \cite{itkndreg-web}, and \sw{PyCA}~\cite{pyca-web}. Literature surveys of image registration can be found in~\cite{Modersitzki:2004a,Sotiras:2013a}. We refer to~\cite{Mang:2018CLAIRE} for a recent overview of existing \lddmm{} methods. Surveys of GPU-accelerated solvers for image registration are~\cite{Fluck:2011a,Shams:2010a,Eklund:2013a}; particular examples for various formulations are~\cite{Budelmann:2019a,Bone:2018b,Courty:2008a,Durrleman:2014a,Ellingwood:2016a,Gu:2009a,Grzech:2019a,Ha:2009a,Ha:2011a,Joshi:2005a,Koenig:2018a,Modat:2010a,Sommer:2011a,Shackleford:2010a,Shamonin:2014a,ValeroLara:2013a,ValeroLara:2014a}. Multi-GPU implementations for \lddmm{} in the context of atlas construction are described in~\cite{Ha:2009a,Ha:2011a,ValeroLara:2013a,ValeroLara:2014a}. None of the hardware-accelerated \lddmm{} methods cited above, except for \claire{}~\cite{Brunn:2020a,Gholami:2017SC,Mang:2015NK,Mang:2016H1,Mang:2016SC,Mang:2018CLAIRE,claire-web}, use second-order information for numerical optimization. Many of the available methods reduce the number of unknowns by using coarser resolutions either through parameterization or by solving the problem on coarser grids; they use simplified algorithms and deliver subpar registration quality.

The work most pertinent to ours is~\cite{Brunn:2020a,Ha:2009a,Ha:2011a}. In~\cite{Ha:2009a,Ha:2011a}, a multi-node multi-GPU implementation of the algorithm in~\cite{Joshi:2005a} is presented. The considered application is atlas construction from multiple image volumes. While computational throughput on a single GPU is optimized, the focus is on data-parallelism: Multiple input images are loaded and synchronously processed on distinct GPUs. We propose a multi-node multi-GPU framework with high computational throughput for single (large-scale) registration problems. This problem is no longer embarrassingly parallel. The computational bottlenecks in~\cite{Ha:2009a,Ha:2011a,Joshi:2005a} are the repeated solution of a Helmholtz-type PDE and trilinear scattered data interpolation to apply the deformation map. The PDE is solved via an implicit successive over-relaxation method. The trilinear interpolation kernel is hardware accelerated with 3D texture volume support. The runtime for a single dataset of size $160\times192\times160$ is \SI{20}{\second} on an NVIDIA Quadro FX 5600. The work in~\cite{Brunn:2020a} presents a single-node single-GPU implementation of \claire{}. The present work ports \claire{} to a heterogeneous multi-node multi-GPU environment by exploiting CUDA-aware MPI. We present several improvements over the computational kernels described in~\cite{Brunn:2020a} (see contributions above).

\section{Discretization and Numerical Algorithms}\label{s:numerics}
To solve~\eqref{e:problem}, we apply the method of Lagrange multipliers to obtain the Lagrangian functional,
\[
\begin{aligned}
\fun{L}(\vect{\phi}) \defeq &
\half{1}\int_{\Omega}\!(\sta(\x,1) - \sta_1(\x))^2\dx
+\frac{\beta}{2} \operatorname{reg}(\vel) \\
& + \int_0^1\!\!\int_\Omega\adj(\x,t) (\p_t \sta + \vel \cdot \igrad \sta )\dx\dt \\
&
+ \int_\Omega\adj(\x,0) (\sta(\x,0) - \sta_0(\x)) \dx,
\end{aligned}
\]
\noindent with state, adjoint, and control variables $(\sta,\adj,\vel)\defeq \vect{\phi}$, respectively.

\paragraph{\textbf{Optimality Conditions \& Reduced Space Approach}}
We derive first-order optimality conditions by taking variations with respect to the state variable $\sta$, the adjoint variable $\adj$, and the control variable $\vel$. This results in a set of coupled, hyperbolic-elliptic PDEs in \emph{4D (space-time)}, consisting of three equations. At optimality, we require that the gradient of our problem vanishes. \claire{} uses a \ipoint{reduced-space approach}, in which one iterates only on the reduced-space of $\vel$. We require that $\vect{g}(\vel) = \vect{0}$, where
\begin{equation}
\label{e:reducedgrad}
\vect{g}(\vel) \defeq \beta \dop{A}\vel(\x) + \int_0^1 \!\!\!\adj(\x,t)\igrad\sta(\x,t)\dt.
\end{equation}

\noindent is the so-called \ipoint{reduced gradient} system (variation of $\fun{L}$ with respect to $\vel$). To evaluate \eqref{e:reducedgrad}, we first solve the forward problem~\eqref{e:varopt:constraint} (variation of $\fun{L}$ with respect to $\adj$) and then the \emph{adjoint problem} (variation of $\fun{L}$ with respect to $\sta$)
\begin{align}
-\p_t\adj(\x,t) - \idiv \adj(\x,t)\vel(\x)  &= 0 && \text{in } \Omega\times[0,1)
\label{e:adj-transport}
\end{align}

\noindent with final condition $\adj(\x,t) =  m_1(\x) - m(\x,t)$ in $\Omega\times\{1\}$ and periodic boundary conditions on $\p\Omega$. \claire{} uses a Newton--Krylov method to solve the non-linear problem $\vect{g}(\vel) = \vect{0}$ as described below.

\paragraph{\textbf{Discretization}}
The forward and adjoint PDEs in the space-time interval $\Omega\times[0,1]$, $\Omega \defeq [0,2\pi)^3\subset\ns{R}^3$, with periodic boundary conditions on $\p\Omega$, are discretized on a regular grid with $N = N_1N_2N_3$ grid points $\x_{ijk}\in\ns{R}^3$ in space and $N_t+1$ grid points in time. A semi-Lagrangian scheme is used to solve the transport equations that appear in the optimality system~\cite{Mang:2017SL,Mang:2016SC}. That is, the advection term is discretized in space and time based on backward trajectories of grid points. The total time derivative is evaluated by means of the difference of the current value of the transported variable at a grid point and the previous time step's value at the end point of a backward trajectory in time. An interpolation in space is needed at the end points of the backward trajectories that are, in general, off-grid points. The backward trajectories themselves are calculated by solving an ODE of the form $\p_t \traj(t) = \vel(\traj(t))$ in $[t,t+\delta t)$ with final condition $\traj(t+\delta t)=\x$ using a second-order Runge--Kutta scheme.

Aside from integrating the PDEs in time, we need to apply gradient and divergence operators to evaluate $\vect{g}$ in \eqref{e:reducedgrad} and to solve~\eqref{e:adj-transport} for $\lambda$. \claire{} uses finite difference (\acr{FD}) operators for these differential operators~\cite{Brunn:2020a}. The reduced gradient~\eqref{e:reducedgrad} also involves the vector-Laplacian $\dop{A}$ and a Leray(-type) projection (see \cite{Mang:2016H1}). These operators are implemented in the spectral domain since \bipa\item as we will see below, the solver requires the application of the inverse of $\dop{A}$ and \item the Leray projection also involves the inverse of a Laplacian operator\eipa. In spectral methods, inverting higher order differential operators can be done at the cost of two FFTs and one a Hadamard product. Using a different scheme would introduce significant complications.

\paragraph{\textbf{Gauss--Newton--Krylov Solver}}
\claire{} uses a Gauss--Newton--Krylov method globalized with an Armijo line search. The iterative scheme is given by
\begin{equation}
\label{e:iter}
\dvel_{k+1} = \dvel_k + \alpha_k \divel_k,
\quad \di{H}\divel_k = -\di{g}_k,
\quad k = 0,1,2,\ldots
\end{equation}

\noindent where $\di{H}\in\ns{R}^{3N,3N}$ is the discretized reduced-space Hessian operator, $\divel_k\in\ns{R}^{3N}$ the search direction, $\di{g}_k\in\ns{R}^{3N}$ a discrete version of the gradient in~\eqref{e:reducedgrad}, $\alpha_k>0$ a line search parameter, and $k\in\ns{N}$ the Gauss--Newton iteration index. We have to solve the linear system in~\eqref{e:iter} at each Gauss--Newton step. We do not form or assemble $\di{H}$; we use a matrix-free preconditioned conjugate gradient (\acr{PCG}) method. This only requires an expression for applying the Hessian matrix to a vector (\ipoint{Hessian matvec}). In the continuum, the Gauss--Newton approximation of this matvec is given by
\begin{equation}
\label{e:matvec}
\dop{H}\ivel = \beta\dop{A}\ivel(\x) + \int_0^1\iadj(\x,t)\igrad \sta(\x,t) \d t.
\end{equation}

\noindent To evaluate this matvec we have to find $\tilde{\lambda}$. Likewise to evaluating the gradient in~\eqref{e:reducedgrad}, this necessitates the solution of two PDEs backward and forward in time, namely
\begin{equation}
\label{e:incsta}
\partial_t \ista(\x,t) + \vel(\x)\cdot\nabla\ista(\x,t) + \ivel(\x)\cdot\nabla \sta(\x,t) = 0
\end{equation}

\noindent in $\Omega \times (0,1]$ and
\begin{equation}
\label{e:incadj}
-\partial_t \iadj(\x,t) - \idiv \iadj(\x,t)\vel(\x) = 0
\quad \text{in }\Omega\times[0,1)
\end{equation}

\noindent with initial and final conditions $\tilde{m}(\x,t) = 0$ in $\Omega \times\{0\}$ and $\iadj(\x,t) = -\ista(\x,t)$ in $\Omega \times \{1\}$, respectively. Inverting $\di{H}$ in \eqref{e:iter} is the most expensive part of \claire{}. We propose a new preconditioner for the reduced-space system in \eqref{e:iter} to amortize the computational costs.

\paragraph{\textbf{Preconditioning}}
As we can see in~\eqref{e:matvec}, the Hessian operator consists of two terms. In a discrete setting, we have $\di{H} = \di{A} + \di{\tilde{H}}$. Here, $\di{A} \in\ns{R}^{3N,3N}$ corresponds to the regularization operator $\dop{A}$ and forming $\di{\tilde{H}} \in\ns{R}^{3N,3N}$ involves $3N$ solutions of~\eqref{e:incsta} and~\eqref{e:incadj}. Given that each Hessian matvec involves two PDE solves, we have to keep the number of PCG iterations as small as possible. With this in mind, we propose a new preconditioner.

As a benchmark, we consider a spectral preconditioner \pcIR{} based on the inverse of $\di{A}$---a common choice in PDE-constrained optimization~\cite{Alexanderian:2016a,BuiThanh:2013a,Mang:2018a} and the default option in \claire{}~\cite{Mang:2015NK,Mang:2016SC,Gholami:2017SC}. This preconditioner is given by
\begin{equation}
\di{s} = (\beta\di{A})^{-1}\di{r},
\end{equation}

\noindent where $\di{r}$ is the residual of the Krylov solver. The cost of applying $(\beta\di{A})^{-1}$ to a vector is two FFTs and a Hadamard product in spectral space.

The proposed preconditioner is based on a zero-velocity approximation of $\di{H}$. This allows us to evaluate the Hessian matvec without having to solve~\eqref{e:incsta} and/or~\eqref{e:incadj}. We term this preconditioner \pcIHN{}. For $\dvel = \vect{0}$, the reduced-space Hessian system in~\eqref{e:iter} becomes $\di{H}_0\divel = -\di{g}$, where $\di{H}_0 \defeq (\beta\di{A} + \nabla \di{m}_0 \otimes \nabla \di{m}_0)$. Here, $\di{m}_0$ is a discrete representation of the template image and $\otimes$ denotes the outer product. It is important to notice that $\di{m}_0$ does not change during the course of the iterations. We use an (approximate) inverse of $\di{H}_0$ as a preconditioner. To compute the action of $\di{H}_0^{-1}$ we iteratively solve the linear system
\begin{equation}
\label{e:h0-pc}
(\beta\di{A} + \nabla \di{m}_0 \otimes \nabla \di{m}_0)\di{s} = \di{r}
\end{equation}

\noindent using a matrix-free PCG method with a relative tolerance $\epsilon_{\dop{H}_0}\epsilon_K$. Here, $\epsilon_K > 0$ is the tolerance for the outer PCG and $\epsilon_{\dop{H}_0} \in (0,1)$. (We need to use a smaller tolerance in the inner PCG since the preconditioner would not act as a linear operator otherwise. We set $\epsilon_{\dop{H}_0}$ to $\num{1e-3}$ for the NIREP data and to $\num{1e-2}$ for the CLARITY data for our runs (see \tabref{t:clairebrain}). These values were determined by experimentation in an attempt to obtain optimal runtimes per type of dataset.)

To compute the inverse of $\di{H}_0$ efficiently, we propose several twists. First, we left-precondition $\di{H}_0$ in \eqref{e:h0-pc} with $(\beta\di{A})^{-1}$ (this adds vanishing computational costs; see above). Second, since $\di{H}_0$ represents a zero-velocity approximation to $\di{H}$, we expect the performance of the preconditioner to deteriorate as we iterate. As a remedy, we replace $\di{m}_0$ in~\eqref{e:h0-pc} with the deformed template image obtained for the current iterate $\dvel_k$ at the beginning of each Gauss-Newton iteration. Third, to further amortize the computational costs, we consider a second variant of \pcIHN{} that exploits a coarse grid discretization. We term this variant \pcIHNTL{}. Here, we invert $\di{H}_0$ on a coarse grid with half the resolution of the fine grid. We restrict the residual $\di{r}$ and $\igrad \di{m}_0$ in $\eqref{e:h0-pc}$. The restriction and prolongation operators are implemented in the spectral domain. \pcIHNTL{} operates only on the low frequency components of $\di{r}$. The solution of the iterative solver, $\di{s}_c$, found on the coarse grid is prolonged to the fine grid and added to the filtered high frequency part of the original residual on the fine grid. In this context, the left-preconditioner $(\beta\di{A})^{-1}$ can be viewed as a (poor) approximation of a multi-grid smoother. Algorithm~\ref{alg:PC} gives an overview of the two proposed preconditioner variants.

{
\begin{center}
\vspace{0.2cm}
\begin{minipage}[c]{0.75\textwidth}
\begin{algorithm}[H]
\caption{Algorithmic overview of the two variants of the \pcIHN{} preconditioner.\label{alg:PC}}
\small
\SetKwFor{Run}{run}{}{}
\SetKwFor{Func}{func}{}{}
\SetKwFor{Block}{}{}{}
\SetKw{Line}{\ }
\Func{$\textsc{InvH0PC}(\di{r})$}{
\Line{$\di{s}_{f}\gets(\beta\di{A})^{-1}\di{r}$, \quad $\text{tol} \gets \epsilon_{\dop{H}_0}\epsilon_K$}\\
\Line{$\di{s}_f\gets$\ \Run(\algcomment{solve \eqref{e:h0-pc}}){$\textsc{CG}(\di{H}_0,\di{s}_f, (\beta\di{A})^{-1}\!\!, \text{tol})$}{
}}
\Return{$\di{s}_f$}
}\vspace*{1em}
\Func{$\textsc{TwoLvlInvH0PC}(\mathbf{r})$}{
\Line{$\di{s}_{f}\gets(\beta\di{A})^{-1}\mathbf{r}$, \quad$\text{tol} \gets \epsilon_{\dop{H}_0}\epsilon_K$}\\
\Line{$\di{s}_{c}\gets\textsc{Restrict}(\di{s}_{f})$}\\
\Line{$\di{s}_c\gets$\ \Run(\algcomment{solve \eqref{e:h0-pc} on coarse grid}){$\textsc{CG}(\di{H}_{0,c},\di{s}_c,(\beta\di{A})^{-1}\!\!,  \text{tol})$}{
}}
\Line{$\di{s}_f\gets\textsc{Prolong}(\di{s}_c) + \textsc{HighPass}(\di{s}_f)$}\\
\Return{$\di{s}_f$}
}
\end{algorithm}
\end{minipage}
\vspace{0.2cm}
\end{center}
\par
}

We observed that the performance of \pcIHN{} deteriorates for vanishing $\beta$. We found by experimentation that, if we use a lower bound of $\num{5e-2}$ for $\beta$ in~\eqref{e:h0-pc}, the preconditioner remains effective even for vanishing $\beta$s for the overall problem. That is, if $\beta < \num{5e-2}$, we set $\beta$ in~\eqref{e:h0-pc} to $\num{5e-2}$.

Finally, the suggested setting for \claire{} is to use a $\beta$-continuation scheme for the solution of the inverse problem~\eqref{e:problem}~\cite{Mang:2015NK,Mang:2018CLAIRE,Brunn:2020a}. That is, \claire{} solves the registration problem for a vanishing sequence of values for $\beta$. For each new value, the velocity obtained at the former step is used as an initial guess for the Gauss-Newton-Krylov solver. For large $\beta$, the problem is dominated by the regularization operator $\di{A}$. As a consequence, the problem is not only easy to solve but the spectral preconditioner is also quite effective. Therefore, if \claire{} is executed using a $\beta$-continuation scheme we use \pcIR{} for $\beta > \num{5e-1}$ and switch to either variant, \pcIHN{} or \pcIHNTL{}, for $\beta \leq \num{5e-1}$ (this bound has been determined by experimentation).

\section{Computational Kernels}\label{s:kernel}
In this section, we describe the multi-node multi-GPU implementation of our computational kernels. In \algref{alg:GNK}, we summarize the overall algorithm. We identify the three most important kernels and their overall contribution to the computational cost: interpolation (\acr{IP}), finite differences (\acr{FD}), and fast-Fourier transforms (\acr{FFT}s). The costs of solving $\di{g}(\dvel) = \di{0}$ (first-order optimality conditions, where $\di{g}$ is a discrete version of \eqref{e:reducedgrad}) for $\dvel(\x)$ are
\begin{equation}
\label{e:totalcosts}
c_{\text{total}} \approx n_{\text{GN}}
\left( n_{\text{CG}}
\left(2 c_{\text{PDE}} + c_{\di{H}} + c_{\text{PC}} \right)
+ 2 c_{\text{PDE}}\right),
\end{equation}

\noindent where $n_{\text{GN}}$ is the number of Gauss--Newton iterations, $n_{\text{CG}}( 2 c_{\text{PDE}} + c_{\di{H}} + c_{\text{PC}})$ summarizes the cost of computing the Gauss-Newton step in~\eqref{e:iter}, $n_{\text{CG}}$ is the number of PCG iterations per Gauss-Newton step (assuming that it is constant to simplify the analysis). The cost for evaluating~\eqref{e:matvec} is denoted by $c_{\di{H}}$. The cost $c_{\text{PC}}$ is for the application of the preconditioner (e.g., iteratively solving \eqref{e:h0-pc}). $c_{\text{PDE}}$ is a prototypical cost for solving the forward or adjoint equations; in particular,~\eqref{e:incsta} and~\eqref{e:incadj}. Let $c_{\text{FD}}$ denote the cost for the FD gradient and $c_{\text{IP}}$ the cost for evaluating the IP kernel for a scalar field, then $c_{\text{PDE}}$ for the RK2 implementation of the semi-Lagrangian scheme is $\mathcal{O}(N_t (c_{\text{FD}} + 4c_{\text{IP}}))$ for~\eqref{e:incsta} (if we choose to not store the gradient of the state variable during the solution of~\eqref{e:varopt:constraint}) and $\mathcal{O}(N_tc_{\text{IP}})$ for~\eqref{e:incadj}. The remaining $2 c_{\text{PDE}}$ in~\eqref{e:totalcosts} are for evaluating the objective functional~\eqref{e:problem} (which involves the solution of~\eqref{e:varopt:constraint}) and the solution of the adjoint problem in~\eqref{e:adj-transport}. The cost $c_{\di{H}}$ for evaluating~\eqref{e:matvec} is dominated by $2 c_{\text{FFT}}$ for applying the regularization operator in the spectral domain (or its inverse) and $N_t c_{\text{FD}}$ (if we choose to not store the gradient of the state variable). The cost for the preconditioner $c_{\text{PC}}$ depends on the choice of the preconditioner. That is, $c_{\text{PC}}$ is $\mathcal{O}(2c_{\text{FFT}})$ for \pcIR{},  $\mathcal{O}(2 c_{\text{FFT}} n_{\text{CG,PC}})$ for \pcIHN{}, and $\mathcal{O}\left(2c_{\text{FFT}} \frac{1}{8} \left( 2 c_{\text{FFT}} n_{\text{CG,PC}} \right)\right)$ for \pcIHNTL, where $n_{\text{CG, PC}}$ is the number of PCG iterations to compute the action of the inverse of $\di{H}_0$. (We kept some of the constant factors to explicitly document the computational steps.). The computational and communication components of $c_{\text{IP}}$, $c_{\text{FD}}$ and $c_{\text{FFT}}$ are reported in~\secref{s:int},~\secref{s:fdm} and~\secref{s:fft}, respectively. We refer to~\cite{Brunn:2020a}, where a DRAM based (ignoring cache heirarchy) roofline analysis is performed for the IP and FD kernels (on a single GPU). DRAM memory accesses for each kernel are modelled analytically assuming full reuse. The number of floating point operations are also estimated analytically. The arithmetic intensity, which is defined as the ratio of total number of floating point operations to number of bytes accessed, is assessed based on this model. The analytical value is compared with the experimental value obtained by the NVIDIA profiler. It is found that both kernels are bound by the GPU DRAM bandwidth.

The work in~\cite{Brunn:2020a} discusses several technical optimizations beyond a pure transition to GPUs, in particular, several options for the IPs as the most important kernel in the semi-Lagrangian solver. In addition, \cite{Brunn:2020a} suggests to replace FFTs used in~\cite{Mang:2018CLAIRE} for first order derivatives by FD approximations. In~\cite{Brunn:2020a}, it is shown empirically that this does not deteriorate the accuracy if FD kernels of high enough order are used. In the following, we describe the implementation of different variants of these kernels, which includes optimizations compared to the work in~\cite{Brunn:2020a} for efficient execution on a multi-node multi-GPU architecture.

{
\begin{center}
\vspace{0.2cm}
\begin{minipage}[c]{0.6\textwidth}
\begin{algorithm}[H]
\caption{Overview of the Gauss--Newton--Krylov solver implemented in \claire{}.}
\label{alg:GNK}
\small
\SetKwFor{Run}{run}{}{}
\SetKwFor{Block}{}{}{}
\SetKw{Line}{\ }
\Line{$\dvel\gets \di{v}_{\text{init}}$, \quad $\epsilon_N \gets \num{5e-2}$}\\
\ \Run{$\textsc{NewtonSolver}(\dvel, \epsilon_N)$}{\vspace*{1mm}
\Line{$\di{m} \gets \textsc{solStateEQ}(\dvel, \di{m}_0)$}\algcomment{solve~\eqref{e:varopt:constraint}}\\
\Line{$\di{\lambda} \gets \textsc{solAdjointEQ}(\dvel, \di{m}, \di{m}_1)$}\algcomment{solve~\eqref{e:adj-transport}}\\
\Line{$\di{g} \gets \textsc{evalGrad}(\dvel, \di{m}, \di{\lambda})$}\algcomment{evaluate~\eqref{e:reducedgrad}}\\
\Line{$\epsilon_K \gets \min(\sqrt{\|\di{g}\|_{\text{rel}}}, 0.5)$}\\
\ \Run(\algcomment{solve~\eqref{e:iter}}){$\textsc{PCG}(\textsc{MatVec}, -\di{g}, \epsilon_{K})$}{
\Block{$\textsc{MatVec}(\divel)$}{
\Line{$\tilde{\di{m}} \gets \textsc{solIncStateEQ}(\dvel, \divel, \di{m})$}\algcomment{solve~\eqref{e:incsta}}\\
\Line{$\tilde{\di{\lambda}} \gets \textsc{solIncAdjointEQ}(\dvel, \tilde m)$}\algcomment{solve~\eqref{e:incadj}}\\
\Line{$\di{H}\divel \gets \textsc{evalMatVec}(\divel, \di{m}, \tilde{\di{\lambda}})$}\algcomment{eval~\eqref{e:matvec}}\\
\Line{$\di{r} \gets -\di{g} - \di{H}\divel$} \\
}
\Block{$\textsc{applyPrecond}(\mathbf{r})$}{see~\algref{alg:PC} \\
}
}
\Run{$\textsc{LineSearch}(\alpha)$}{
\Line{$m \gets \textsc{solStateEQ}(\dvel+\alpha\divel, \di{m}_0)$}\algcomment{solve~\eqref{e:varopt:constraint}}\\
\Line{$\textsc{evalObjective}(\dvel+\alpha\divel, \di{m})$}\algcomment{eval~\eqref{e:varopt:objective}}\\
}
\Line{$\dvel \gets \dvel + \alpha \divel$}\algcomment{Newton step}
}
\end{algorithm}
\end{minipage}
\vspace{0.2cm}
\end{center}
\par
}

The total memory consumption mostly depends on the domain size $N = N_1N_2N_3$. The state variable $\sta(\x, t)$ has to be stored for all time steps to avoid additional PDE solves. The memory footprint for the proposed method is
\[
\begin{aligned}
\mu_{\text{total}} & \approx \mu_\text{PDE} + \mu_\text{FFT} + \mu_\text{FD} + \mu_\text{SL} + \mu_\text{GN/CG} + \mu_\text{IP} + \mu_\text{API}\nonumber\\
 &= ((24 + N_t) + 7 + 2 + 11 + 30)\nicefrac{N \mu_0}{p} + \mu_\text{IP} + \mu_\text{API}\nonumber\\
 &= (74 + N_t)\nicefrac{N \mu_0}{p} + \mu_\text{IP} + \mu_\text{API},
\end{aligned}
\]

\noindent where $\mu_0$ is word size of the datatype (i.e. 4\,byte for single precision floating point values). The memory required for the ghost layer communication in the IP model is $\mu_{\text{IP}} \approx 30 d N_2 N_3 \mu_0$ with polynomial degree $d$. Note that the runtime API overhead, $\mu_{\text{API}}$, depends on $N$ (especially for \texttt{cuFFT}~\cite{Nvidia2007b} and \texttt{PETSc}~\cite{Balay:2020a,petsc-web}), but is not further estimated.

\subsection{Interpolation}\label{s:int}
The semi-Lagrangian scheme requires IP of vector and scalar fields along backward characteristics. We use Lagrange polynomial-based cubic IP but also consider first-order trilinear IP since GPUs offer hardware acceleration through texture units (not fully single-precision). The formula for interpolating at an off-grid query point $\x = (x_1,x_2,x_3)$ is given by
\[
\textstyle f(\x)
= \sum_{i,j,k=0}^{d}f_{ijk}\phi_i(x_1)\phi_j(x_2)\phi_k(x_3),
\]

\noindent where $f_{ijk}$ is the function value at a grid point, $d$ is the polynomial order and $\phi_l$, $l=0,\ldots,d$, are the Lagrange polynomial basis functions. The numerical accuracy and compute performance of variants of the IP kernel on a single GPU have been discussed in~\cite{Brunn:2020a}. We focus on optimizations for the multi-GPU implementation. We follow the workflow described in~\cite{Mang:2016SC,Gholami:2017SC} with the following \ipoint{major modifications}:
\begin{enumerate}
\item[\textbf{1)}]
We use CUDA-aware MPI to reduce or eliminate expensive on-node host-device transfers.

\item[\textbf{2)}]
We use the \texttt{thrust} library~\cite{Thrust} to efficiently determine, which query points need to be processed by which GPU, thereby completely eliminating host-side computation.

\item[\textbf{3)}]
We use a sparse point-to-point communication to send points on the backward characteristics to other processors, as proposed in~\cite{Gholami:2017SC}. We adaptively allocate memory for the respective MPI send and receive buffers using an estimate of the maximal displacement of grid points along backward trajectories based on the CFL number of the velocity field.

\item[\textbf{4)}]
Following~\cite{Brunn:2020a}, we perform local IP on a single GPU using GPU-TXTLAG or GPU-TXTLIN (for high-resolution images). Although GPU-TXTSPL in~\cite{Brunn:2020a} is much faster than GPU-TXTLAG on a single GPU, for the distributed memory implementation it requires ghost layer communication for the pre-filtering step, which makes it slower than GPU-TXTLAG.
\end{enumerate}

The computational cost $c_{\text{IP}}$ of applying the IP kernel GPU-TXTLAG is $\mathcal{O}(\nicefrac{482N}{p})$ (see~\cite{Brunn:2020a}), where $p$ is the number of processors and $N=N_1N_2N_3$. For GPU-TXTLIN, it is $\mathcal{O}(\nicefrac{30N}{p})$. The total cost of communicating ghost points, query points and interpolated values is $\mathcal{O}(u_{\max}N_2N_3)$ where $u_{\max}\in\ns{R}$ is an estimate of the maximum displacement of a voxel from a regular grid point along the coordinate directions. For the IP kernel we do not consider overlapping communication and computation because of the data dependencies in the semi-Lagrangian scheme.

We perform a weak scaling experiment for an isolated semi-Lagrangian solve on a real dataset and present the runtime breakdown in~\tabref{tab:weak_interp_breakdown}. We use a realistic velocity field for this experiment (obtained by registration of two brain images) to ensure a representative scenario for the communication of query points between MPI ranks. The \ipoint{major observations} are:
\begin{enumerate}
\item[\textbf{1)}]
Since we use slab decomposition in $x_1$-dimension, the message size for \texttt{ghost\_comm} is $\mathcal{O}(N_2N_3)$. Hence, it roughly doubles every time $N_2$ or $N_3$ is doubled.

\item[\textbf{2)}]
We see a similar increase for \texttt{interp\_comm} and \texttt{scatter\_comm}. Due to the non-uniformity in space of the query points, communication time does not double exactly and we observe an imbalance in the communication for different MPI ranks.

\item[\textbf{3)}]
The time spent in \texttt{interp\_kernel} is almost the same across all cases and takes up the majority of the time for up to 16 GPUs. Beyond 16 GPUs, communication dominates the overall runtime.

\item[\textbf{4)}]
Since we are performing scattered IP, determining which and how many query points need to be processed locally or sent to other MPI ranks in \texttt{scatter\_mpi\_buffer} leads to expensive scattered memory accesses.\footnote{We rely on the \texttt{thrust::copy\_if} algorithm for this purpose.} This explains why \texttt{scatter\_mpi\_buffer} requires almost one third of \texttt{interp\_kernel} runtime.
\end{enumerate}

\begin{table*}
\caption{Weak scaling study for the IP kernel. We report runtimes for the semi-Lagrangian scheme. We advect a real brain MRI (\texttt{na10} of the NIREP data; see~\secref{s:results}) with a velocity field obtained from the registration of \texttt{na10} to \texttt{na01}. We use cubic IP GPU-TXTLAG and $N_t = 4$ time steps. We report the runtime (in seconds) of the major components in the algorithm and their percentages with respect to the total runtime. These components are \texttt{ghost\_comm} (communication of ghost points), \texttt{interp\_comm} (communication of interpolated values), \texttt{scatter\_comm} (communication of query points), \texttt{interp\_kernel} (IP kernel), \texttt{scatter\_mpi\_buffer} (creation of MPI buffer for sending query points to other ranks). The experiments were performed on TACC's Longhorn system with four Nvidia V100 GPUs per node and a single GPU per MPI rank. We scale from a single GPU to 64 GPUs for grid resolutions ranging from $256^3$ to $1024^3$.\label{tab:weak_interp_breakdown}}
\centering
\resizebox{\textwidth}{!}{%
\begin{tabular}{lrRrRrRrRrRrRrR}
\toprule
size
& \multicolumn{2}{r}{$256\PLH256\PLH256$}
& \multicolumn{2}{r}{$512\PLH256\PLH256$}
& \multicolumn{2}{r}{$512\PLH512\PLH256$}
& \multicolumn{2}{r}{$512\PLH512\PLH512$}
& \multicolumn{2}{r}{$1024\PLH512\PLH512$}
& \multicolumn{2}{r}{$1024\PLH1024\PLH512$}
& \multicolumn{2}{r}{$1024\PLH1024\PLH1024$} \\ \midrule
	\#GPUs &        1 &      \% &        2 &      \% &        4 &      \% &        8 &      \% &       16 &      \% &       32 &      \% &       64 &      \% \\
\midrule
\texttt{ghost\_comm}          & \num{0.0}      &   0.0 & \num{2.48E-03} &   7.6 & \num{3.49E-03} &   9.9 & \num{7.51E-03} &  18.0 & \num{8.66E-03} &  19.1 & \num{1.31E-02} &  24.0 & \num{2.23E-02} &  31.3 \\
\texttt{interp\_comm}         & \num{0.0}      &   0.0 & \num{1.71E-03} &   5.2 & \num{1.80E-03} &   5.1 & \num{3.62E-03} &   8.7 & \num{4.17E-03} &   9.2 & \num{5.92E-03} &  10.9 & \num{9.73E-03} &  13.6 \\
\texttt{scatter\_comm}        & \num{0.0}      &   0.0 & \num{2.65E-04} &   0.8 & \num{7.81E-04} &   2.2 & \num{2.02E-03} &   4.8 & \num{2.85E-03} &   6.3 & \num{5.42E-03} &  10.0 & \num{8.72E-03} &  12.2 \\ \midrule
\texttt{interp\_kernel}       & \num{1.77E-02} &  93.3 & \num{1.79E-02} &  54.8 & \num{1.76E-02} &  49.8 & \num{1.76E-02} &  42.0 & \num{1.83E-02} &  40.2 & \num{1.84E-02} &  33.9 & \num{1.87E-02} &  26.2 \\ \midrule
\texttt{scatter\_mpi\_buffer} & \num{0.00E+00} &   0.0 & \num{5.88E-03} &  18.0 & \num{7.16E-03} &  20.3 & \num{6.63E-03} &  15.9 & \num{6.98E-03} &  15.4 & \num{7.00E-03} &  12.9 & \num{7.30E-03} &  10.2 \\ \midrule
\textbf{total}                & \num{1.90E-02} & 100.0 & \num{3.28E-02} & 100.0 & \num{3.53E-02} & 100.0 & \num{4.18E-02} & 100.0 & \num{4.54E-02} & 100.0 & \num{5.44E-02} & 100.0 & \num{7.13E-02} & 100.0 \\
\bottomrule
\end{tabular}}
\end{table*}

\subsection{Finite Differences} \label{s:fdm}
The CPU version of \claire{} uses FFTs for spatial derivatives~\cite{Mang:2016SC,Gholami:2017SC,Mang:2018CLAIRE}. Since our functions are periodic, these spectral operators are diagonal. \cite{Brunn:2020a} proposes a mixed-accuracy implementation that replaces the spectral discretization of the divergence and gradient operators with a FD scheme. This mixed scheme is more accurate (for the considered grid sizes---not asymptotically) and faster than differentiation via FFTs. In particular, an $8^{\text{th}}$ order central difference scheme is used. We extend the single-GPU FD kernel described in~\cite{Brunn:2020a} to a multi-node multi-GPU environment. The computational cost $c_{\text{FD}}$ of applying the FD kernel is $\mathcal{O}(\nicefrac{20N}{p})$, where $p$ is the number of processors and $N=N_1N_2N_3$. To compute derivatives at the boundary of our 2D slab decomposition, we communicate a ghost layer of size $\mathcal{O}(N_2N_3)$ to neighboring MPI ranks. We perform strong and weak scaling experiments for computing the gradient of a synthetic scalar field; see~\tabref{tab:fd8}. For a single GPU, no communication is involved. It is much faster than using multiple GPUs (for small problem sizes). In the weak scaling setup, the runtime increases when we switch from one to eight to 64 GPUs because the size of the ghost layer increases ($N_2$ and $N_3$ increase), while the kernel execution time itself remains constant. In the strong scaling setting, the kernel scales well for up to 8 GPUs. Beyond 8 GPUs, the kernel execution time becomes much smaller than the communication time (which is constant); this negatively impacts the scalability. Since the FD kernel is not a bottleneck---as seen in~\tabref{tab:clairescaling}---we did not explore the idea of overlapping communication and computation when evaluating the kernel.

\begin{table}
\caption{Scalability for our finite difference (FD) scheme for first order derivatives. We show strong scaling for $512^3$ from one to eight MPI ranks, and weak scaling for $256^3$ to $1024^3$ from one to 64 MPI ranks. We report the breakdown of runtime (in seconds) into \texttt{comm} (communication of ghost points) and \texttt{kernel} (FD kernel) and show percentages with respect to total runtime.\label{tab:fd8}}
\scriptsize\centering
\begin{tabular}{rrrRrRr}
\toprule
\#GPUs & size  & \texttt{comm} &   \% & \texttt{kernel} &    \% & \textbf{total} \\\midrule
 1 & $256^3$   &          0.0  &  0.0 &   \num{6.32E-4} & 100.0 &  \num{6.32E-4} \\
 1 & $512^3$   &          0.0  &  0.0 &   \num{4.82E-3} & 100.0 &  \num{4.82E-3} \\
 2 & $512^3$   & \num{9.37E-4} & 21.9 &   \num{3.33E-3} &  78.1 &  \num{4.27E-3} \\
 4 & $512^3$   & \num{7.01E-4} & 29.2 &   \num{1.70E-3} &  70.8 &  \num{2.40E-3} \\
 8 & $512^3$   & \num{9.86E-4} & 53.2 &   \num{8.66E-4} &  46.8 &  \num{1.85E-3} \\
16 & $512^3$   & \num{8.94E-4} & 66.0 &   \num{4.60E-4} &  34.0 &  \num{1.35E-3} \\
64 & $1024^3$  & \num{2.85E-3} & 76.0 &   \num{9.03E-4} &  24.0 &  \num{3.76E-3} \\
\bottomrule
\end{tabular}
\end{table}

\subsection{FFT}\label{s:fft}
The distributed memory implementation of \claire{}~\cite{Gholami:2017SC,Mang:2016SC,Mang:2018CLAIRE} uses \texttt{AccFFT} \cite{accfft_github, accfft-home-page}, which supports MPI for CPUs and GPUs. In~\cite{Brunn:2020a}, \texttt{cuFFT}~\cite{Nvidia2007b} is used, as they focus on a single-GPU implementation. Higher order derivatives and their inverses require 3D FFTs. \texttt{AccFFT} uses a pencil decomposition (see, e.g., \cite{Mang:2016SC}), which is efficient for 1D FFTs (needed for divergence and gradient operators). In \cite{Brunn:2020a}, $1^{\text{st}}$ order derivatives have been replaced by FD kernels. For the proposed multi-GPU implementation, we use a combination of \texttt{cuFFT} and a new 2D slab decomposition, which allows us to use the highly optimized 2D \texttt{cuFFT} on each GPU. We decompose the spatial domain in the outer-most dimension (i.e., $x_1$) and in the spectral domain in $x_2$ dimension. Thus, the inner-most $x_3$ dimension is always continuous in memory. This reduces misaligned memory accesses for communication and transpose operations. The real-to-complex transformation is divided into three steps. \bipa\item We use \texttt{cuFFT}'s batched 2D FFTs in the $x_2$--$x_3$ plane. \item The complex data are transposed to a decomposition in $x_2$ dimension. \item We apply \texttt{cuFFT}'s batched 1D FFTs to the $x_1$ dimension, which is non-continuous in memory\eipa. For the inverse complex-to-real transformation, these three steps are executed in reverse order, using the respective inverse transformations. The complexity for communication of the 2D slab decomposition is $\mathcal{O}(\nicefrac{N}{P} - \nicefrac{N}{P^2})$ per process. If the FFT is executed on a single rank, we still use \texttt{cuFFT}'s 3D FFT to avoid additional operations, in particular an explicit transpose operation on the data and misaligned memory accesses. Also, it reduces the number of memory accesses of the spectral data from device memory.

For communication between GPUs, we use CUDA-aware MPI. We found that \texttt{MPI\_Alltoallv} (IBM Spectrum MPI 10.3~\cite{ibmspectrum-web}) is not optimized for direct GPU communication. For communication volumes larger than $\sim$\SI{500}{\kilo\byte}, all-to-all communication using direct GPU-optimized peer-to-peer routines is faster on our test system (see~\tabref{tab:MPI_alltoall}). We implement a threshold of \SI{512}{\kilo\byte} to switch between an asynchronous peer-to-peer communication scheme or \texttt{MPI\_Alltoallv}. For FFTs on a single node (four GPUs), we always use the peer-to-peer scheme to utilize the \emph{NVLink} inter-GPU bus. The communication is only overlapped with the process-local transpose operation due to data dependencies.

\begin{table}
\begin{minipage}{0.5\textwidth}
\caption{MPI performance analysis for the proposed FFT kernel. We use one GPU per MPI rank. We report the sustained bidirectional CUDA MPI bandwidth in GB/s. The results are averaged over ten runs and the smallest value for all ranks is presented. We compare \texttt{MPI\_Alltoall} to our own implementation using asynchronous peer-to-peer routines. The local data size per rank is $8N_1N_2(\lfloor N_3/2 \rfloor + 1)/p$ byte. The peer-to-peer communication volume is $8N_1N_2(\lfloor N_3/2 \rfloor + 1)/p^2$ byte. Runs in the shaded cells have a communication volume larger than \SI{512}{\kilo\byte}. The fastest runs are highlighted in bold.\label{tab:MPI_alltoall}}
\end{minipage}
\begin{minipage}{0.5\textwidth}
\scriptsize\centering
\begin{tabular}{rrrrrrrr}\toprule
\multicolumn{2}{c}{\textbf{setup}} & \multicolumn{6}{c}{\textbf{MPI tasks}}\\
\cmidrule(lr){1-2}\cmidrule(lr){3-8}
size            & type & 4                                   & 8                                   & 16  & 32  & 64  & 128\\\hline
$256^3$         & MPI  & \cellcolor{lightgray}          5.6  & \cellcolor{lightgray}           5.0 &                       \textbf{3.3} &                       \textbf{2.2} &                       \textbf{2.0} & \textbf{1.5} \\
                & P2P  & \cellcolor{lightgray} \textbf{35.7} & \cellcolor{lightgray} \textbf{ 9.3} &                                2.2 &                                1.3 &                                1.6 &          1.4 \\ \hdashline
$512\PLH256^2$  & MPI  & \cellcolor{lightgray}          5.1  & \cellcolor{lightgray}           5.2 &                                3.5 &                       \textbf{1.5} &                       \textbf{1.9} & \textbf{1.9} \\
                & P2P  & \cellcolor{lightgray} \textbf{36.0} & \cellcolor{lightgray} \textbf{ 9.5} &                       \textbf{5.8} &                                1.0 &                                1.5 &          1.4 \\ \hdashline
$512^2\PLH256$  & MPI  & \cellcolor{lightgray}          5.4  & \cellcolor{lightgray}          4.6  & \cellcolor{lightgray}          3.5 &                       \textbf{2.8} &                                1.6 & \textbf{2.7} \\
                & P2P  & \cellcolor{lightgray} \textbf{36.6} & \cellcolor{lightgray} \textbf{ 9.9} & \cellcolor{lightgray} \textbf{6.1} &                                0.4 &                       \textbf{1.7} &          1.4 \\ \hdashline
$512^3$         & MPI  & \cellcolor{lightgray}          5.9  & \cellcolor{lightgray}          4.9  & \cellcolor{lightgray}          3.9 & \cellcolor{lightgray}          2.7 &                       \textbf{2.5} & \textbf{2.7} \\
                & P2P  & \cellcolor{lightgray} \textbf{37.1} & \cellcolor{lightgray} \textbf{ 9.5} & \cellcolor{lightgray} \textbf{5.9} & \cellcolor{lightgray} \textbf{4.7} &                                0.5 &          1.5 \\ \hdashline
$1024\PLH512^2$ & MPI  & \cellcolor{lightgray}          6.4  & \cellcolor{lightgray}          5.4  & \cellcolor{lightgray}          3.9 & \cellcolor{lightgray}          3.4 &                       \textbf{3.2} & \textbf{2.2} \\
                & P2P  & \cellcolor{lightgray} \textbf{32.6} & \cellcolor{lightgray} \textbf{10.1} & \cellcolor{lightgray} \textbf{5.9} & \cellcolor{lightgray} \textbf{4.8} &                                0.4 &          0.5 \\\hdashline
$1024^2\PLH512$ & MPI  & \cellcolor{lightgray}           6.7 & \cellcolor{lightgray}          5.5  & \cellcolor{lightgray}          4.2 & \cellcolor{lightgray}          3.6 & \cellcolor{lightgray}          3.4 & \textbf{2.7} \\
                & P2P  & \cellcolor{lightgray} \textbf{36.6} & \cellcolor{lightgray}\textbf{10.5}  & \cellcolor{lightgray} \textbf{5.4} & \cellcolor{lightgray} \textbf{4.7} & \cellcolor{lightgray} \textbf{4.5} &          0.3 \\\hdashline
$1024^3$        & MPI  & \cellcolor{lightgray}           6.7 & \cellcolor{lightgray}           5.6 & \cellcolor{lightgray}          4.4 & \cellcolor{lightgray}          3.7 & \cellcolor{lightgray}          3.4 & \textbf{3.1} \\
                & P2P  & \cellcolor{lightgray} \textbf{36.8} & \cellcolor{lightgray} \textbf{10.6} & \cellcolor{lightgray} \textbf{5.2} & \cellcolor{lightgray} \textbf{4.6} & \cellcolor{lightgray} \textbf{4.3} &          0.4 \\\bottomrule
\end{tabular}
\end{minipage}
\end{table}

\begin{table}
\caption{Weak (diagonals) and strong (rows) scaling for the proposed 3D FFT kernel in slab decomposition (forward and inverse). We use one GPU per MPI rank. We report the runtime in ms. The FFT uses CUDA-aware MPI. We switch from point-to-point communication to \texttt{MPI\_Alltoall} for small slabs. Results are averaged over 20 runs. For a single rank, the runtime is also given for \texttt{cuFFT} 3D-FFTs (3D). The highlighted runs use peer-to-peer communication.\label{tab:FFT}}
\scriptsize\centering
\begin{tabular}{rrrrrrrrr}\toprule
& & \multicolumn{7}{c}{\textbf{MPI tasks}}\\\cmidrule(lr){3-9}
\textbf{size}    &   3D &    1 &             4 &             8 &            16 &            32 &            64 & 128  \\\hline\rowcolor{lightgray}
$256^3$          & 1.41 & 1.86 & \textbf{2.83} & \textbf{3.92} & 4.17          &          3.88 &          2.93 & 3.76 \\
$512\PLH256^2$   & 3.20 & 3.87 & \textbf{5.39} & \textbf{7.65} & \textbf{7.33} &          5.21 &          4.09 & 4.30 \\
$512^2\PLH256$   & 7.30 & 7.70 & \textbf{8.48} & \textbf{13.8} & \textbf{13.3} &          8.29 &          5.67 & 5.12 \\\rowcolor{lightgray}
$512^3$          & 16.9 & 16.9 & \textbf{15.6} & \textbf{25.7} & \textbf{24.5} & \textbf{16.7} &          9.63 & 7.23 \\
$1024\PLH512^2$  & 31.2 & 40.1 & \textbf{31.8} & \textbf{51.3} & \textbf{43.6} & \textbf{31.3} &          17.8 & 11.8 \\
$1024^2\PLH512$  &  --- &  --- & \textbf{65.7} & \textbf{100}  & \textbf{90.5} & \textbf{54.2} & \textbf{33.4} & 21.4 \\\rowcolor{lightgray}
$1024^3$         &  --- &  --- & \textbf{132}  & \textbf{198}  & \textbf{182}  & \textbf{116}  & \textbf{62.0} & 38.4 \\\bottomrule
\end{tabular}
\end{table}

In addition to the memory footprint of \texttt{cuFFT} our 2D slab decomposition needs twice the local domain size to execute an out-of-place transformation. The temporary memory consumtion of \texttt{cuFFT} is between $\nicefrac{2N}{p}$ and $\nicefrac{16N}{p}$ real valued elements~\cite{Nvidia2007b}. \tabref{tab:FFT} shows that our 3D FFT with 2D slab decomposition is almost as fast as \texttt{cuFFT} 3D-FFT, but can be accelerated and scaled to data sizes beyond the memory capacity of a single GPU. Given the $\mathcal{O}(N \log N)$ computational complexity of the FFT (with data size $N$) and the huge amount of data communication inherent to FFTs, we observe good scalability up to 128\,GPUs, for the large problem sizes---even in strong scaling.

\section{Results}\label{s:results}
We \bipa\item analyze the numerical and runtime efficiency of our new preconditioner and \item assess the overall scalability and efficiency of our multi-GPU multi-node implementation\eipa.

We use the following datasets:
\begin{enumerate}
\item[\textbf{1)}]
\textbf{SYN} is a synthetic test problem, where the template image is $m_0(\x) \defeq \sum_{i=1}^3\sin^2(x_i)/3$ and the reference image $m_1(\x)$ is computed by solving~\eqref{e:varopt:constraint} with initial condition $m_0(\x)$ and given velocity $\vel(\x) \defeq (\sin(x_i),\cos(x_k),\sin(x_k))_{(i,k) = (3,2), (1,3), (2,1)}$.

\item[\textbf{2)}]
\textbf{NIREP}~\cite{Christensen:2006a} is a standardized repository for assessing registration accuracy that contains 16 T1-weighted MR neuroimaging datasets (\texttt{na01}--\texttt{na16}) of different individuals (see \figref{f:regprob-brains}). The original image size is $256\PLH300\PLH256$ voxels.

\item[\textbf{3)}]
\textbf{CLARITY}~\cite{Chung:2013b,Kim:2013a,Kutten:2016a,Kutten:2017a,Tomer:2014a,Vogelstein:2018a,clarity-web} are biomedical imaging datasets with a resolution of $\SI{0.6}{\micro\metre}\PLH\SI{0.6}{\micro\metre}\PLH\SI{6}{\micro\metre}$ and a grid size at the order of $20\text{\,K}\PLH20\text{\,K}\PLH1\text{\,K}$ (see \figref{f:regprob-clarity}) . We have affinely pre-registered these datasets (at a much lower resolution) using \sw{FAIR}~\cite{Modersitzki:2009a} prior to executing \claire{}.
\end{enumerate}

All runs were executed on \ipoint{TACC's Longhorn system} in single precision. Longhorn hosts 96 NVIDIA Tesla V100 nodes. Each node is equipped with four GPUs with $4\PLH16$\,GB GPU RAM (64\,GB aggregate) and two IBM Power 9 processors with 20 cores (40 cores per node) at 2.3\,GHz with 256\,GB memory. Our implementation uses \texttt{PETSc}~\cite{Balay:2020a,petsc-web} for linear algebra, \texttt{PETSc}'s \texttt{TAO} package for the nonlinear optimization, \texttt{CUDA}~\cite{cuda-web}, \texttt{thrust}~\cite{Thrust}, \texttt{cuFFT} for FFTs~\cite{Nvidia2007b}, \texttt{niftilib}~\cite{niftilib-web} for I/O, IBM Spectrum MPI~\cite{ibmspectrum-web}, and the IBM XL compiler~\cite{ibmxl-web}.

\subsection{Preconditioning}\label{sec:res_precond}
We study different preconditioner variants. We use the datasets \texttt{na02}, \texttt{na03}, and \texttt{na10} from the NIREP repository as template images, and \texttt{na01} as reference image.

\begin{figure}
\begin{minipage}{0.6\textwidth}
\includegraphics[width=9cm]{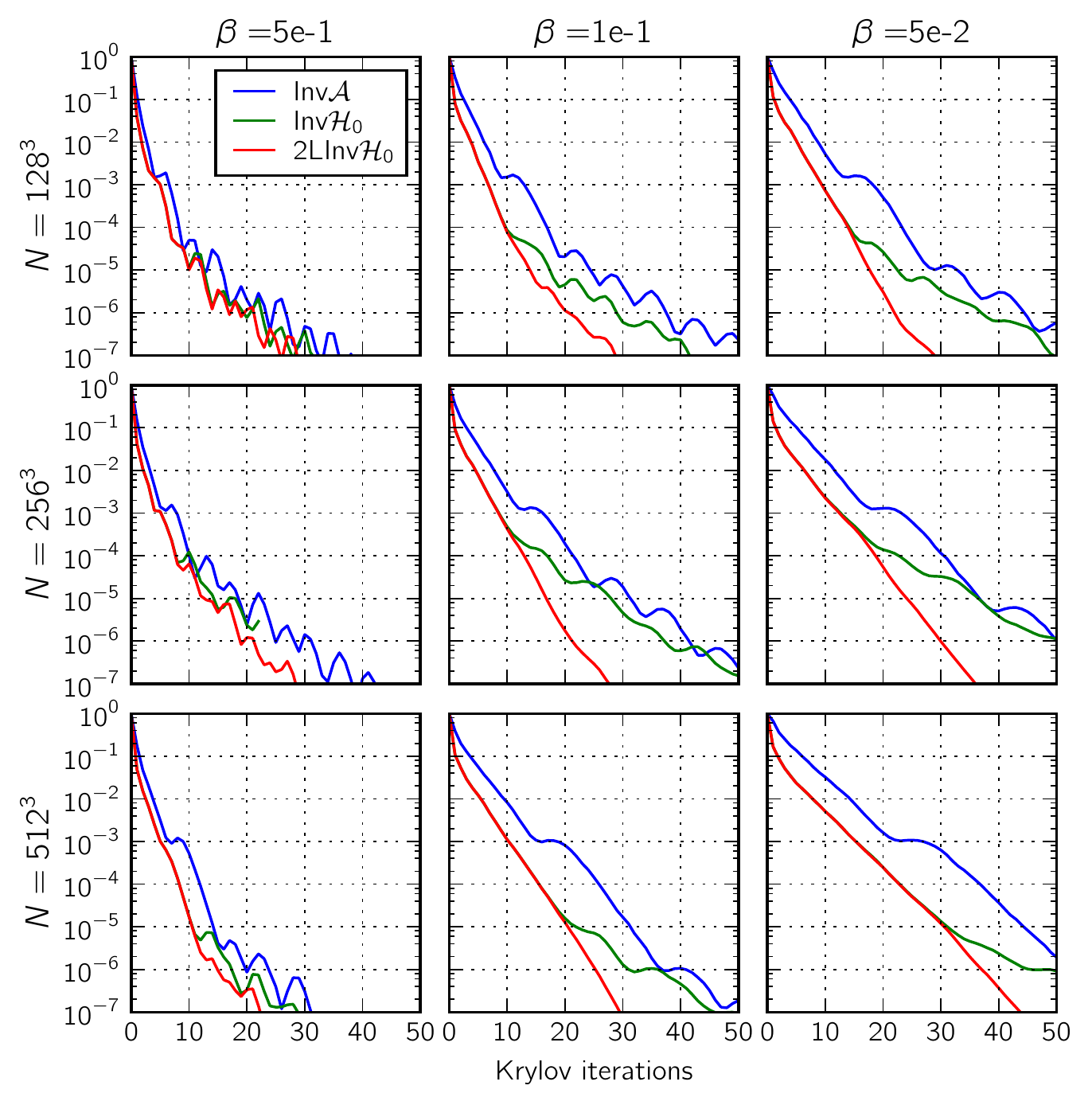}
\end{minipage}
\begin{minipage}{0.38\textwidth}
\caption{We report the trend of the PCG residual versus PCG iterations for the benchmark preconditioner \pcIR{} used in~\cite{Mang:2018CLAIRE,Brunn:2020a} and the proposed preconditioner variants \pcIHN{} and \pcIHNTL{}. We vary the regularization parameter $\beta$ (columns; $\beta\in\{\num{5e-1}, \num{1e-1},\num{5e-2}\}$) and the domain size $N$ (rows; $N\in\{128^2, 256^3,512^3\}$). We solve the problem at the true solution (see text for a description).\label{fig:h0residual}}
\end{minipage}
\end{figure}

\paragraph{\textbf{Results}} We report convergence plots for a single Gauss-Newton step in~\figref{fig:h0residual}. We initialize the solver with \texttt{na10} as template and a reference image synthetically generated by solving the forward problem with a true registration velocity (\texttt{na10} to \texttt{na01}). The true (non-zero) velocity is used as an initial guess for the Gauss-Newton-Krylov method (i.e., we solve \eqref{e:iter} at the solution of the inverse problem). This allows us to assess \bipa\item the convergence at a point in the optimization landscape at which we expect the PCG to take many iterations and \item identify potential issues that may arise due to a zero-velocity approximation at a point at which the velocity is non-zero.\eipa We report results for varying grid sizes and values for $\beta$.

\paragraph{\textbf{Observations}} The proposed preconditioner leads to faster convergence (fewer iterations) and is less sensitive to a reduction in $\beta$ than \pcIR. We expect the preconditioner to be mesh-independent but not $\beta$-independent. All preconditioners exhibit (close to) mesh independent behavior. Interestingly, for the considered range for $\beta$, \pcIHNTL{} is close to being $\beta$-independent; only for $\beta=\num{5e-2}$ we see the performance slightly deteriorate as the mesh size increases. In general, we expect that we might have to use larger values for $\beta$ for higher resolutions, since higher frequencies can occur in the images and the velocity field (coarsening can be viewed as an additional regularization).

\subsection{Registration Performance}\label{sec:regres}
We study the performance of the proposed methods for the solution of the inverse registration problem. We report results for three different template images from the NIREP repository: \texttt{na02}, \texttt{na03}, and \texttt{na10}. For \texttt{na10}, we increase the resolution from $256^3$ to $1024^3$ (spectral prolongation). Results for the registration of the dataset \texttt{na10} to \texttt{na01} are shown in \figref{f:regprob-brains}. We expect the convergence behavior of the Gauss-Newton-Krylov method to be independent of the mesh size. In addition to that, we report results for the registration of two representative CLARITY volumes (dataset Cocaine 175 to Control 189; Control 189 is visualized in \figref{f:regprob-clarity}). We consider all preconditioner variants.

\begin{table*}
\caption{Results for the registration of different NIREP and CLARITY datasets. We report results for the different preconditioners \pcIR{} (\textbf{[A]}), \pcIHN (\textbf{[B]}), and \pcIHNTL{} (\textbf{[C]}). We use a parameter continuation scheme for $\beta$ with target parameter $\beta = \num{5e-4}$. The number of time steps for the semi-Lagrangian method is $N_t = 4, 8, 16$ for domain sizes $N = 256^3,512^3,1024^3$, respectively. All runs use linear IP and FD for $1^{\text{st}}$ order derivatives. For each domain size, we use the minimum number of resources possible, i.e., a single GPU for $N = 256^3$, four GPUs on a single node for $N = 512^3$, 32 GPUs on 8 nodes for $N= 1024^3$. We report from left to right: (\textbf{data}) the selected template image, (\textbf{PC}) the Hessian preconditioner method, (\textbf{GN}) the number of Gauss-Newton  iterations, (\textbf{PCG}) the number of PCG iterations, (\textbf{mism.}) the relative mismatch, ($||\di{g}||_{\text{rel}}$) the relative gradient norm,  (\textbf{[A]}) the number of applications of \pcIR{}, (\textbf{[B|C]}) the number of applications of \pcIHN/\pcIHNTL (notice, that we use \pcIR{} for large values of $\beta$ in our continuation scheme), (\textbf{total}) and (\textbf{average}) the number of PCG iterations to invert $\di{H}_0$ (total and average), (\textbf{PC}) the overall runtime for preconditioner, (\textbf{Obj}) objective function evaluation, (\textbf{Grad}) reduced gradient computation, (\textbf{Hess}) Hessian matvecs, and (\textbf{Total}) the total runtime of the entire solver. All runtimes are in seconds.\label{t:clairebrain}}
\scriptsize\centering
\begin{tabular}{rrrrccrrrrllll>{\columncolor{lightgray}}ll}\toprule
\multicolumn{2}{c}{\textbf{setting}}&\multicolumn{4}{c}{\textbf{solver}}&\multicolumn{4}{c}{\textbf{preconditioner}}&\multicolumn{5}{c}{\textbf{runtimes}}\\
\cmidrule(lr){1-2}\cmidrule(lr){3-6}\cmidrule(lr){7-10}\cmidrule(lr){11-15}
&&\multicolumn{2}{c}{iterations}&\multicolumn{2}{c}{relative accuracy}&\multicolumn{2}{c}{applications}&\multicolumn{2}{c}{CG steps}\\
data&PC& GN&PCG&mism.&$||\di{g}||_{\text{rel}}$&A&B|C& total & avg. & PC&Obj&Grad&Hess&Total\\
\hline
\rowcolor{colorA}\multicolumn{15}{l}{\textbf{NIREP} $N=256^3$, $N_t=4$, $\epsilon_{\mathcal{H}0}=$1e-3, 1 node, 1 GPU}\\\hline
na02 & [A] & 14 &  75 & \num{2.73e-2} & \num{3.09e-2} &  75 & --- &  --- &  --- & \num{4.43e-1} & \num{2.04e-1} & \num{4.33e-1} & \num{3.82e0} & \num{6.19e0}\\
     & [B] & 14 &  23 & \num{2.62e-2} & \num{2.82e-2} &   3 &  20 &  235 & 11.8 & \num{2.45e0 } & \num{2.04e-1} & \num{4.33e-1} & \num{1.27e0} & \num{5.54e0}\\
     & [C] & 14 &  28 & \num{2.79e-2} & \num{3.23e-2} &   3 &  25 &  294 & 11.8 & \num{1.04e0 } & \num{2.05e-1} & \num{4.35e-1} & \num{1.52e0} & \num{4.44e0}\\
\hdashline
na03 & [A] & 17 &  93 & \num{2.55e-2} & \num{3.11e-2} &  93 & --- &  --- &  --- & \num{5.50e-1} & \num{2.49e-1} & \num{5.24e-1} & \num{4.69e0} & \num{7.53e0}\\
     & [B] & 17 &  36 & \num{2.50e-2} & \num{3.04e-2} &  14 &  22 &  255 & 11.6 & \num{2.72e0 } & \num{2.48e-1} & \num{5.23e-1} & \num{1.91e0} & \num{6.80e0}\\
     & [C] & 17 &  39 & \num{2.56e-2} & \num{3.17e-2} &  14 &  25 &  301 & 12.0 & \num{1.11e0 } & \num{2.49e-1} & \num{5.24e-1} & \num{2.05e0} & \num{5.39e0}\\
\hdashline
na10 & [A] & 17 &  94 & \num{1.96e-2} & \num{2.94e-2} &  94 & --- &  --- &  --- & \num{5.58e-1} & \num{2.50e-1} & \num{5.25e-1} & \num{4.76e0} & \num{7.61e0}\\
     & [B] & 17 &  36 & \num{1.90e-2} & \num{2.81e-2} &   9 &  27 &  299 & 11.1 & \num{3.17e0 } & \num{2.48e-1} & \num{5.25e-1} & \num{1.91e0} & \num{7.25e0}\\
     & [C] & 17 &  38 & \num{1.93e-2} & \num{2.90e-2} &   9 &  29 &  328 & 11.3 & \num{1.22e0 } & \num{2.49e-1} & \num{5.26e-1} & \num{2.01e0} & \num{5.45e0}\\
\hline
\rowcolor{colorA}\multicolumn{15}{l}{\textbf{NIREP} $N=512^3$, $N_t=8$, $\epsilon_{\mathcal{H}0}=$1e-3, 1 node, 4 GPUs}\\\hline
na10 & [A] & 18 & 107 & \num{2.53e-2} & \num{3.84e-2} & 107 & --- &  --- &  --- &  \num{5.28e0} & \num{1.68e0} & \num{3.86e0} & \num{3.52e1} & \num{5.18e1}\\
     & [B] & 18 &  37 & \num{2.66e-2} & \num{4.38e-2} &  10 &  27 &  307 & 11.4 & \num{2.19e1} & \num{1.70e0} & \num{3.89e0} & \num{1.25e1} & \num{4.55e1}\\
     & [C] & 18 &  37 & \num{2.68e-2} & \num{4.39e-2} &  10 &  27 &  309 & 11.4 & \num{5.55e0} & \num{1.67e0} & \num{3.87e0} & \num{1.25e1} & \num{2.92e1}\\
\hline
\rowcolor{colorA}\multicolumn{15}{l}{\textbf{NIREP} $N=1024^3$, $N_t=16$, $\epsilon_{\mathcal{H}0}=$1e-3, 8 nodes, 32 GPUs}\\\hline
na10 & [A] & 21 & 128 & \num{3.19e-2} & \num{4.41e-2} & 128 & --- &  --- &  --- & \num{4.63e1} & \num{3.55e0} & \num{2.14e1} & \num{1.76e2} & \num{2.55e2}\\
     & [B] & 22 &  59 & \num{2.70e-2} & \num{3.34e-2} &  18 &  41 &  531 & 13.0 & \num{2.33e2} & \num{3.79e0} & \num{2.24e1} & \num{8.08e1} & \num{3.46e2}\\
     & [C] & 22 &  59 & \num{2.73e-2} & \num{3.77e-2} &  18 &  41 &  533 & 13.0 & \num{5.69e1} & \num{3.80e0} & \num{2.24e1} & \num{8.11e1} & \num{1.71e2}\\
\hline
\rowcolor{colorA}\multicolumn{15}{l}{\textbf{CLARITY} $N=1024\PLH384\PLH384$, $N_t=4$, $\epsilon_{\mathcal{H}0}=$1e-2, 1 nodes, 4 GPUs}\\\hline
     & [A] & 13 & 205 & \num{2.01e-1} & \num{4.23e-2} & 205 & --- &  --- &  --- & \num{2.12e1} & \num{8.78e-1} & \num{2.53e0} & \num{5.11e1} & \num{7.13e1}\\
     & [C] & 12 &  75 & \num{2.02e-1} & \num{4.54e-2} &   4 &  71 & 1007 & 14.2 & \num{1.67e1} & \num{8.49e-1} & \num{2.34e0} & \num{1.89e1} & \num{4.36e1}\\
\hline
\rowcolor{colorA}\multicolumn{15}{l}{\textbf{CLARITY} $N=1024\PLH768\PLH768$, $N_t=4$, $\epsilon_{\mathcal{H}0}=$1e-2, 4 nodes, 16 GPUs}\\\hline
     & [A] & 20 & 663 & \num{1.95e-1} & \num{5.81e-2} & 663 & --- &  --- &  --- & \num{1.96e2} & \num{4.02e0} & \num{1.37e1} & \num{5.12e2} & \num{7.38e2}\\
     & [B] & 15 &  52 & \num{2.03e-1} & \num{4.38e-2} &   6 &  46 &  648 & 14.1 & \num{2.28e2} & \num{1.57e0} & \num{1.09e1} & \num{4.02e1} & \num{2.86e2}\\
\bottomrule
\end{tabular}
\end{table*}

\begin{figure*}
\centering
\scalebox{0.6}{\input{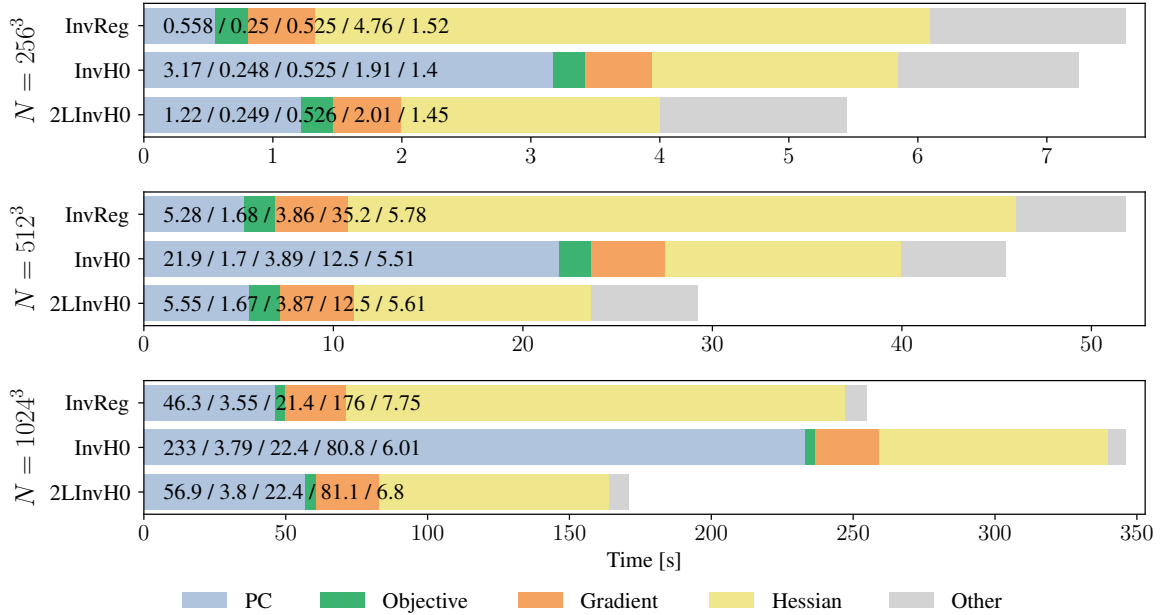}}
\caption{Visualization of the allocated runtime for the results reported in \tabref{t:clairebrain}. The color bars (times are in seconds) illustrate the amount of execution time spent in the main mathematical operators of our solver (PC: application of inverse of preconditioner; objective: evaluation of the objective functional; gradient: evaluation of gradient (includes PDE solves for state and adjoint equation); hessian: Hessian matvecs (includes PDE solves for incremental state and adjoint equation). We can observe that we spend a large fraction of our runtime on the computation of the Newton step.}\label{f:solcomp}
\end{figure*}

\paragraph{\textbf{Results}} The results can be found in~\tabref{t:clairebrain}. We report the number of Gauss-Newton iterations, the accumulated number of PCG iterations across all Gauss-Newton iterations, the relative reduction of the mismatch, the relative reduction of the gradient, the number of applications of the inverse regularization operator, the number of applications of \pcIHN{} or \pcIHNTL{}, the number of PCG iterations to invert $\di{H}_0$ (in total and on average), the time spent in the core parts of the solver, and the total runtime. We visualize the runtime of the solver components in \figref{f:solcomp}.

\begin{figure*}
\centering
\scalebox{0.6}{\input{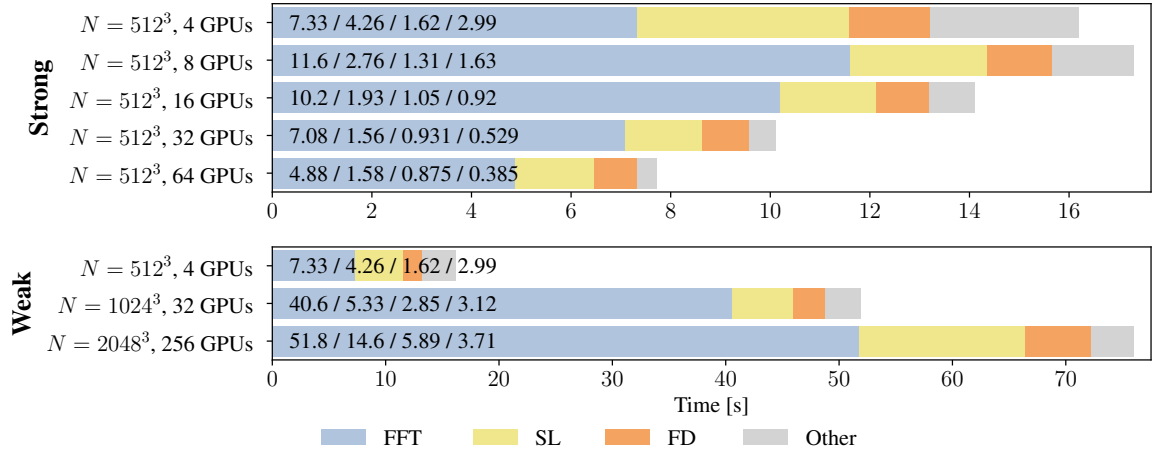}}
\caption{We visualize exemplary strong (top block) and weak (bottom block) scaling results for the experiments reported in \tabref{tab:clairescaling}. For each run, we report the fraction (color bars and runtime in seconds) spent in the individual kernels. We can see that the runtime is dominated by the FFT kernel. We can also observe that almost the entire runtime of our solver is spent in the three main computational kernels---FFTs, SL, and FD. The scalability of our multi-node multi-GPU implementation is limited due to the high communication costs for small local problem sizes and load imbalance across ranks. We provide a more detailed analysis in the text.\label{f:scaling}}
\end{figure*}

\paragraph{\textbf{Observations}} The most important observation is that our solver converges quickly to accurate solutions. We require 14 to 22 Gauss-Newton-Krylov iterations. The number of Gauss-Newton-Krylov and PCG iterations is approximately mesh-independent. The most effective preconditioner is \pcIHNTL{}. If we compare the runtime for our new version to the results reported in~\cite{Brunn:2020a}, we can observe a speedup of about 50\%. The average time-to-solution for clinically relevant problems on a single GPU is \textasciitilde{\SI{5}{\second}}. We can reduce the runtime on a single GPU to \SI{3.7}{\second}, which corresponds to a speedup of 70\% compared to~\cite{Brunn:2020a} (for \texttt{na02}, $256^3$) by storing the gradient of the state variable. Storing the gradient of the state variable reduces the runtime by approximately $15\%$ (but increases the memory pressure). We can also observe that we can solve large-scale real-world imaging problems with grid sizes of $1024^3$ for the NIREP data and up to $1024\PLH768\PLH768$ for the CLARITY data on 8 nodes with 32 GPUs or one 4 nodes with 16 GPUs, respectively. In terms of registration quality, we achieve the same accuracy as reported in~\cite{Brunn:2020a, Mang:2018CLAIRE}. These studies also include comparisons to other LDDMM software packages. They demonstrated that their implementation of \claire{} yields results that are significantly more accurate (in terms of data mismatch) than existing methods, and that the single-node GPU version of \claire{} is up to $30\PLH$ faster than other available single-GPU implementations. With the present work, we are $50\PLH$ faster on a single GPU.

\subsection{Strong and Weak Scaling Results}
We study weak and strong scaling for our new multi-node multi-GPU implementation. We consider the SYN dataset and use the \pcIR{} preconditioner for these runs. We fix the number of Gauss-Newton iterations to 5 and the number of PCG iterations per Newton step to 10 to avoid discrepancies arising from the use of relative tolerances.

\paragraph{\textbf{Results}} We present the results in~\tabref{tab:clairescaling} and report the time-to-solution along with the time spent in individual kernels. We additionally provide the \% of the execution time spent for data communication and the total memory consumption per GPU. The strong and weak scaling experiments are restricted by the slab size and available GPU memory, respectively. For the memory restrictions we refer to the analytical estimates given above. Considering the domain decomposition, we cannot use arbitrarily many GPUs per problem size since the slab size (local data volume) per GPU becomes too small for the computations to be efficient. We visualize strong scaling for $N=512^3$ and weak scaling in \figref{f:scaling}.

\paragraph{\textbf{Observations}} The most important observations are \bipa\item we can solve problems of unprecedented scale (the $1024^3$ and the $2048^3$ problem can not be solved on a single GPU; the largest problem solved in~\cite{Brunn:2020a} is $384^3$) and \item the scalability of our solver suffers from high communication costs for small local problem sizes\eipa. In particular, the runtime in FFTs is dominated by communication because of the required all-to-all collective. For a single GPU, we utilize the \texttt{cuFFT} 3D FFT and need no additional memory transfers. For small problem sizes (e.g., $128^3$ or $256^3$), the additional communication costs for strong scaling cannot be compensated by the reduced computations per rank. For all FFTs, scaling above a single node (4 ranks) increases the runtime due to off-node communication, which is the limiting factor. In~\tabref{tab:weak_interp_breakdown}, we considered GPU-TXTLAG to test the scalability of the semi-Lagrangian method. However, here we use GPU-TXTLIN, which has much lower computational complexity. This results in an increased percentage of communication in the overall runtime, and as we reduce the local problem size (slab width $<$16 voxels), this effect is further amplified. At this slab size, the communication of the query points can become non-uniform (subject to local variations in length of the characteristics). This can cause a significant load imbalance among MPI ranks and by that negatively affects the scaling because of the implicit synchronization for the next communication step (which is ghost layer sharing). The scaling performance of the FD kernel is consistent with the results in~\tabref{tab:fd8}. For weak scaling, when switching from $512^3$ on 4 GPUs to $2048^3$ on 256 GPUs, the communication time increases by \approxspeed{4}; the kernel execution time stays roughly the same. For the strong scaling for resolutions $512^3$ and $1024^3$, the communication time stays roughly the same while the kernel execution time reduces by \approxspeed{2}. However, the overall time spent in FD does not scale well because of GPU memory constraints, as explained in~\secref{s:fdm}.

\begin{table*}
\caption{Strong and weak scaling results for \claire{} using synthetic data. The number of Gauss--Newton iterations is fixed to 5 and we use 10 PCG steps per Gauss-Newton iteration. We consider \pcIR{} as a preconditioner. We report runtimes in seconds and total memory consumtion in GB per GPU. We use a fixed regularization of $\beta=\num{1e-3}$ and set $N_t =4$ with a linear interpolation (IP) model for the semi-Lagrangian (SL) method. All $1^{\text{st}}$ order derivatives are computed with FDs. The parallel layout (number of GPUs) for our experiments is restricted by the local slab size (can become too small) and the available GPU memory, respectively. The $2048^3$ is the largest problem we could fit on TACC's Longhorn system. We cannot use less resources for this problem due to memory restrictions.\label{tab:clairescaling}}
\centering\vspace*{0.5em}
\footnotesize
\begin{tabular}{rrlrlrlr>{\columncolor{lightgray}}lrr}\toprule
\textbf{nodes} & \#\textbf{GPUs} & \multicolumn{2}{c}{\textbf{FFT}} & \multicolumn{2}{c}{\textbf{SL}} & \multicolumn{2}{c}{\textbf{FD}} & \multicolumn{3}{c}{\textbf{overall}} \\
\cmidrule(lr){3-4}\cmidrule(lr){5-6}\cmidrule(lr){7-8}\cmidrule(lr){9-11} & & time & \% comm. & time & \% comm. & time & \% comm & time & \% comm. & \revise{memory}\\
\hline
\rowcolor{colorA}\multicolumn{11}{l}{$N=128^3$}\\\hline
 1 &   1 & \num{1.03E-01} &   0.0 & \num{1.82E-01} &  0.0 & \num{6.12E-02} &  0.0 & \num{5.11E-01} &  0.0 & \revise{1.11}\\
 1 &   2 & \num{1.74E-01} &  44.5 & \num{3.88E-01} & 69.3 & \num{1.52E-01} & 54.3 & \num{8.37E-01} & 51.3 & \revise{0.95}\\
 1 &   4 & \num{2.35E-01} &  59.8 & \num{4.13E-01} & 76.4 & \num{1.44E-01} & 62.0 & \num{9.17E-01} & 59.5 & \revise{0.79}\\
 2 &   8 & \num{6.95E-01} &  85.5 & \num{5.56E-01} & 83.9 & \num{2.87E-01} & 84.4 & \num{1.66E+00} & 78.4 & \revise{0.71}\\
 4 &  16 & \num{5.38E-01} &  90.0 & \num{6.19E-01} & 85.5 & \num{5.72E-01} & 92.1 & \num{1.87E+00} & 82.3 & \revise{0.66}\\
\hline
\rowcolor{colorA}\multicolumn{11}{l}{$N=256^3$}\\\hline
 1 &   1 & \num{7.74E-01} &   0.0 & \num{1.16E+00} &  0.0 & \num{3.72E-01} &  0.0 & \num{3.32E+00} &  0.0 & \revise{5.09}\\
 1 &   2 & \num{7.47E-01} &  42.3 & \num{1.20E+00} & 61.0 & \num{4.64E-01} & 34.1 & \num{2.99E+00} & 40.5 & \revise{3.18}\\
 1 &   4 & \num{9.84E-01} &  74.7 & \num{8.20E-01} & 66.5 & \num{3.20E-01} & 45.4 & \num{2.56E+00} & 55.6 & \revise{1.95}\\
 2 &   8 & \num{1.69E+00} &  89.2 & \num{1.23E+00} & 85.2 & \num{3.90E-01} & 71.8 & \num{3.60E+00} & 78.9 & \revise{1.29}\\
 4 &  16 & \num{1.96E+00} &  91.8 & \num{1.26E+00} & 89.4 & \num{3.70E-01} & 79.6 & \num{3.81E+00} & 84.5 & \revise{0.94}\\
 8 &  32 & \num{1.36E+00} &  95.3 & \num{1.24E+00} & 91.4 & \num{3.59E-01} & 84.0 & \num{3.15E+00} & 86.8 & \revise{0.78}\\
\hline
\rowcolor{colorA}\multicolumn{11}{l}{$N=512^3$}\\\hline
 1 &   4 & \num{7.33E+00} &  74.0 & \num{4.26E+00} & 60.6 & \num{1.62E+00} & 32.2 & \num{1.62E+01} & 52.5 & \revise{11.2}\\
 2 &   8 & \num{1.16E+01} &  90.0 & \num{2.76E+00} & 68.0 & \num{1.31E+00} & 56.4 & \num{1.73E+01} & 75.5 & \revise{5.84}\\
 4 &  16 & \num{1.02E+01} &  94.5 & \num{1.93E+00} & 74.5 & \num{1.05E+00} & 70.3 & \num{1.41E+01} & 83.9 & \revise{3.32}\\
 8 &  32 & \num{7.08E+00} &  94.3 & \num{1.56E+00} & 81.3 & \num{9.31E-01} & 80.4 & \num{1.01E+01} & 85.9 & \revise{2.00}\\
16 &  64 & \num{4.88E+00} &  96.8 & \num{1.58E+00} & 87.9 & \num{8.75E-01} & 86.9 & \num{7.72E+00} & 89.1 & \revise{1.31}\\
\hline
\rowcolor{colorA}\multicolumn{11}{l}{$N=1024^3$}\\\hline
 8 &  32 & \num{4.06E+01} &  95.0 & \num{5.33E+00} & 73.4 & \num{2.85E+00} & 69.6 & \num{5.19E+01} & 85.7 & \revise{11.5}\\
16 &  64 & \num{2.44E+01} &  95.0 & \num{4.17E+00} & 81.9 & \num{2.48E+00} & 81.4 & \num{3.27E+01} & 87.4 & \revise{6.23}\\
32 & 128 & \num{1.47E+01} &  96.9 & \num{3.94E+00} & 89.2 & \num{2.20E+00} & 88.2 & \num{2.18E+01} & 90.2 & \revise{3.43}\\
64 & 256 & \num{1.00E+01} &  97.5 & \num{6.64E+00} & 96.2 & \num{2.04E+00} & 92.3 & \num{1.95E+01} & 92.9 & \revise{2.12}\\
\hline
\rowcolor{colorA}\multicolumn{11}{l}{$N=2048^3$}\\\hline
64 & 256 & \num{5.18E+01} &  93.1 & \num{1.46E+01} & 92.4 & \num{5.89E+00} & 88.5 & \num{7.60E+01} & 88.1 & \revise{12.5}\\\bottomrule
\end{tabular}
\end{table*}

\section{Conclusions}\label{s:con}
We presented a novel multi-node multi-GPU implementation for diffeomorphic registration. Our work extends the publicly available software package \claire{}. \claire{} relies on three main computational kernels: FFTs and FD kernels for differentiation and the evaluation of IP kernels in a semi-Lagrangian solver for the solution of transport equations. Our approach to port these kernels to a multi-GPU environment is highly adapted to the target architecture in various ways: \bipa\item We replace FFT-based (spectral) first-order derivative evaluations used in \claire{} with an $8^{\text{th}}$ order FD scheme for the multi-GPU version. This yields a scheme that is more accurate (for the considered resolutions and precision; not asymptotically) and, at the same time, requires substantially less communication. (Similar results are reported in \cite{Brunn:2020a} for a single-GPU implementation.) \item We choose texture-based Lagrange polynomial third order IP over spline IP (which had been shown to be superior on a single GPU~\cite{Brunn:2020a}) to further reduce the communication between GPUs. \item We propose an efficient combination of \texttt{cuFFT} within nodes and a 2D slab decomposition approach across nodes, combined with an in-house developed, optimized all-to-all communication for regimes for which we could show that the available vendor MPI all-to-all~\cite{ibmspectrum-web} was sub-optimal\eipa. In addition to these kernel optimizations, we are able to substantially reduce the number of PCG iterations for computing the search direction within a Gauss--Newton--Krylov scheme and, thus, reduce the runtime by a factor of up to 2.5 compared to the prior version of \claire{}. This is achieved through a new two-level (coarse grid) preconditioner based on a zero-velocity approximation of the Hessian operator, which eliminates expensive PDE solves. The entire solver is matrix-free. We optimized the memory footprint of the proposed solver. This allows us to solve larger problems on a single GPU, and to tackle problems of unprecedented scale. We ported \claire{} to multi-GPU architectures as a whole, and support direct GPU-GPU communication through CUDA-aware MPI; no explicit host-to-device communication is required. The largest run reported in this study is $152\PLH$ larger than the results reported for the state-of-the-art~\cite{Brunn:2020a}. Combining all improvements, we achieved a speedup of up to 70\% compared to~\cite{Brunn:2020a} on a single GPU. To showcase the capabilities of the proposed methodology, we reported results for the registration of real imaging data for resolutions of up to $1024^3$ for MR neuroimaging data (on 8 nodes with a total of 32 GPUs) and $1024\PLH768\PLH768$ for CLARITY imaging data (on 4 nodes with a total of 16 GPUs). The achieved accuracy is equivalent to the results provided in prior work on \claire~\cite{Brunn:2020a,Gholami:2017SC,Mang:2016SC,Mang:2018CLAIRE,claire-web}, and on par or superior to other state-of-the-art software for diffeomorphic registration (see~\cite{Brunn:2020a,Mang:2018CLAIRE} for a comparison).

Our work applies to other transport dominated forward and inverse problems. For example, the semi-Lagrangian GPU algorithm applies to particle-in-cell and weather/climate codes. The code basis of our solver (optimization scheme, linear algebra solvers, and preconditioning) are hardware agnostic. Our three main computational kernels should translate to other GPU accelerators as long as they provide some specialized hardware support. For example, the IP kernel relies on texture memory, which needs to be supported by the hardware. Also, certain parameters will need to be retuned. Most of the kernels are written in CUDA, so---although the algorithms won't change---the implementation will have to be ported to the new GPU programming interface.

\subsection*{Acknowledgments} We thank Nicolas Charon and Joshua T. Vogelstein at Johns Hopkins University for assisting us with gaining access to the CLARITY data.

\end{document}